\documentclass{aastex}          
\usepackage{spr-astr-addons}    

\usepackage{float}

\begin{document}
%
\title{Characterising the physical and chemical properties of a young Class~0
protostellar core embedded in the Orion~B9 filament}

\shorttitle{Orion B9--SMM 3}
\shortauthors{Miettinen}

\author{O.~Miettinen\altaffilmark{1,2}} 
\email{oskari@phy.hr} 

\altaffiltext{1}{Department of Physics, University of Zagreb, Bijeni\v{c}ka cesta 32, HR-10000 Zagreb, Croatia}
\altaffiltext{2}{Department of Physics, University of Helsinki, P.O. Box 64, 
FI-00014 Helsinki, Finland}

\begin{abstract}
Deeply embedded low-mass protostars can be used as testbeds to study the 
early formation stages of solar-type stars, and the prevailing chemistry before 
the formation of a planetary system. The present study aims to characterise further the physical and chemical properties of the protostellar core Orion B9--SMM3. The Atacama Pathfinder EXperiment (APEX) telescope was used to perform a follow-up molecular line survey of SMM3. The observations were done using the single pointing (frequency range 218.2--222.2~GHz) and on-the-fly mapping 
methods (215.1--219.1~GHz). These new data were used in conjunction with our 
previous data taken by the APEX and Effelsberg 100~m telescopes. 
The following species were identified from the frequency range 218.2--222.2~GHz: $^{13}$CO, C$^{18}$O, SO, \textit{para}-H$_2$CO, and E$_1$-type CH$_3$OH. The mapping observations revealed that SMM3 is associated with a dense gas core as traced by DCO$^+$ and \textit{p}-H$_2$CO. Altogether three different \textit{p}-H$_2$CO transitions were detected with clearly broadened linewidths ($\Delta v\sim8.2-11$~km~s$^{-1}$ in FWHM). The derived \textit{p}-H$_2$CO rotational temperature, $64\pm15$~K, indicates the presence of warm gas. We also detected a narrow \textit{p}-H$_2$CO line ($\Delta v=0.42$~km~s$^{-1}$) at the systemic velocity. The \textit{p}-H$_2$CO abundance for the broad component appears to be enhanced by two orders of magnitude with respect to the narrow line value ($\sim3\times10^{-9}$ versus $\sim2\times10^{-11}$). The detected methanol line shows a linewidth similar to those of the broad \textit{p}-H$_2$CO lines, which indicates their coexistence. The CO isotopologue data suggest that the CO depletion factor decreases from $\sim27\pm2$ towards the core centre to a value of $\sim8\pm1$ towards the 
core edge. In the latter position, the N$_2$D$^+$/N$_2$H$^+$ ratio is revised 
down to $0.14\pm0.06$. The origin of the subfragments inside the SMM3 core we found previously can be understood in terms of the Jeans instability if non-thermal motions are taken into account.
The estimated fragmentation timescale, and the derived chemical abundances 
suggest that SMM3 is a few times $10^5$~yr old, in good agreement with its 
Class~0 classification inferred from the spectral energy distribution analysis.
The broad \textit{p}-H$_2$CO and CH$_3$OH lines, and the associated warm gas 
provide the first clear evidence of a molecular outflow driven by SMM3. 

\end{abstract}

\keywords{Astrochemistry - Stars: formation - Stars: protostars - ISM: 
individual objects: Orion B9--SMM3}

%

%
%

\section{Introduction}

Low-mass stars have main-sequence masses of $M_{\star}\simeq0.08-2$~M$_{\sun}$, 
and are classified with spectral types of M7--A5 (e.g. \citealp{stahler2005}). 
The formation process of these types of stars begins when the parent molecular 
cloud core undergoes gravitational collapse (e.g. \citealp{shu1987}; 
\citealp{mckee2007}). In the course of time, the collapsing core centre 
heats up due to compression, and ultimately becomes a protostar. 
The youngest low-mass protostars, characterised by accretion from 
the much more massive envelope ($M_{\rm env}\gg M_{\star}$), 
are known as the Class~0 objects (\citealp{andre1993}, 2000).

A curious example of a Class~0 protostellar object is SMM3 in the Orion~B9 
star-forming filament. This object was first uncovered by 
Miettinen et al. (2009; hereafter Paper~I), when they mapped Orion B9 using 
the Large APEX BOlometer CAmera (LABOCA) at 870~$\mu$m. In Paper~I, we constructed and analysed 
a simple mid-infrared--submillimetre spectral energy distribution (SED) of SMM3, and classified it as a Class~0 object. The physical and chemical properties of SMM3 (e.g. the gas temperature and the level of N$_2$H$^+$ deuteration) were further characterised by Miettinen et al. (2010, 2012; hereafter referred to as Papers~II and III, respectively) through molecular line observations. In Paper~III, we also presented the results of our Submillimetre APEX BOlometer 
CAmera (SABOCA) 350~$\mu$m imaging of Orion~B9. With the flux density of 
$S_{350\,{\rm \mu m}}\simeq 5.4$~Jy, SMM3 turned out to be the strongest 
350~$\mu$m emitter in the region. Perhaps more interestingly, the 
350~$\mu$m image revealed that SMM3 hosts two subfragments (dubbed SMM3b and 
3c) on the eastern side of the protostar, where an extension could already 
be seen in the LABOCA map at 870~$\mu$m. The projected distances of the subfragments from the 
protostar's position, 0.07--0.10~pc\footnote{In the present work, we have adopted a distance of $d=420$~pc to the source to be consistent with the most recent studies of SMM3 (\citealp{stutz2013}; \citealp{tobin2015}; \citealp{furlan2016}). We note that in Papers~I--III, we assumed a distance 
of $450$~pc, which is a factor of 1.07 larger than used here.}, 
were found to be comparable to the local thermal Jeans length. 
This led us to suggest that the parent core might have fragmented 
into smaller units via Jeans gravitational instability. 

The Orion~B or L1630 molecular cloud, including Orion~B9, was mapped with 
\textit{Herschel} as part of the \textit{Herschel} Gould Belt Survey (HGBS; Andr\'e et al. 2010)\footnote{The HGBS is a \textit{Herschel} key programme jointly carried out by SPIRE Specialist Astronomy Group 3 (SAG 3), scientists of se\-veral institutes in the PACS Consortium (CEA Saclay, INAF-IFSI Rome and INAF-Arcetri, KU Leuven, MPIA Heidelberg), and scientists of the \textit{Herschel} Science Center (HSC). For more details, see {\tt http://gouldbelt-herschel.cea.fr}}. The \textit{Herschel} images revealed that Orion~B9 is actually a 
filamentary-shaped cloud in which SMM3 is embedded  
(see Fig.~2 in \citealp{miettinen2013b}). Miettinen (2012b)
found that there is a sharp velocity gradient in the parent filament 
(across its short axis), and suggested that it might represent a shock front resulting from the feedback from the nearby expanding H{\scriptsize II} region/OB cluster NGC~2024 ($\sim3.7$~pc to the southwest of Orion~B9). Because SMM3 appears to lie on the border of the velocity gradient, it might have a physical connection to it, and it is possible that the formation of SMM3 (and the other dense cores in Orion~B9) was triggered by external, positive feedback (\citealp{miettinen2012b}). Actually, the OB associations to the west of the whole Orion~B cloud have likely affected much of the cloud area through their strong feedback in the form of ionising radiation and stellar winds (e.g. \citealp{cowie1979}). The column density probability distribution function of Orion~B, studied by Schneider et al. (2013), was indeed found to be broadened as a result of external compression. 

The Class~0 object SMM3 was included in the Orion protostellar core
survey by Stutz et al. (2013, hereafter S13; their source 090003). Using data from 
\textit{Spitzer}, \textit{Herschel}, SABOCA, and LABOCA, S13 
constructed an improved SED of SMM3 compared to what was presented in Paper~I. 
The bolometric temperature and luminosity -- as based on the Myers \& Ladd (1993) method -- were found to be $T_{\rm bol}=36.0\pm0.8$~K and $L_{\rm bol}=2.71\pm0.24$~L$_{\sun}$. They also performed a modified blackbody (MBB) fit to the SED ($\lambda \geq70$~$\mu$m) of SMM3, and obtained a dust temperature of $T_{\rm dust}=21.4\pm0.4$~K, luminosity of $L=2.06\pm0.15$~L$_{\sun}$, 
and envelope mass of $M_{\rm env}=0.33\pm0.06$~M$_{\sun}$ (see their Fig.~9). The derived SED properties led S13 to the conclusion that SMM3 is likely a Class~0 object, which supports our earlier suggestion (Papers~I and III). 

Tobin et al. (2015) included SMM3 in their Combined Array for Research for Millimetre Astronomy (CARMA) 2.9~mm continuum imaging survey of Class~0 objects in Orion. This was the first high angular resolution study of SMM3. With a 2.9~mm flux density of $S_{\rm 2.9\, mm}=115.4\pm3.9$~mJy (at an angular resolution of $2\farcs74 \times 2\farcs56$), SMM3 was found to be the second brightest source among the 14 target sources. The total (gas$+$dust) mass 
derived for SMM3 by Tobin et al. (2015), $M=7.0\pm0.7$~M$_{\sun}$,  is much higher than that derived earlier by S13 using a MBB fitting technique, which underpredicted the 870~$\mu$m flux density of the source (see \citealp{tobin2015} and Sect.~4.1 herein for further discussion and different assumptions used). Tobin et al. (2015) did not detect 2.9~mm emission from the subfragments SMM3b or 3c, which led the authors to conclude that they are starless. 

Kang et al. (2015) carried out a survey of H$_2$CO and HDCO emission towards Class~0 objects in Orion, and SMM3 was part of their source sample (source HOPS403 therein). The authors derived a HDCO/H$_2$CO ratio of $0.31\pm0.06$ for SMM3, which improves our knowledge of the chemical characteristics of this source, and strongly points towards its early evolutionary stage from a chemical point of view. 

Finally, we note that SMM3 was part of the recent large Orion protostellar core survey by Furlan et al. (2016; source HOPS400 therein), where the authors presented the sources' panchromatic (1.2--870~$\mu$m) SEDs and radiative transfer model fits. They derived a bolometric luminosity of $L_{\rm bol}=2.94$~L$_{\sun}$ (a trapezoidal summation over all the available flux density data points), total (stellar$+$accretion) luminosity of $L_{\rm tot}=5.2$~L$_{\sun}$, bolometric temperature of $T_{\rm bol}=35$~K (following \citealp{myers1993} as in S13), and an envelope mass of $M_{\rm env}=0.30$~M$_{\sun}$, which are in fairly good agreement with the earlier S13 results. We note that the total luminosity derived by Furlan et al. (2016) from their best-fit model is corrected for inclination effects, and hence is higher than $L_{\rm bol}$. Moreover, the aforementioned value of $M_{\rm env}$ refers to a radius of 2\,500~AU ($=0.012$~pc), which corresponds to an angular radius of about $6\arcsec$ at the distance of SMM3, while a similar envelope mass value derived by S13 refers to a larger angular scale as a result of coarser resolution of the observational data used (e.g. $19\arcsec$ resolution in their LABOCA data).
 
In the present study, we attempt to further add to our understanding of the physical and chemical 
properties of SMM3 by means of our new molecular line observations. We also 
re-analyse our previous spectral line data of SMM3 in a uniform manner to 
make them better comparable with each other. This paper is outlined as 
follows. The observations and the observational data are described 
in Sect.~2. The immediate observational results are presented in Sect.~3. 
The analysis of the observations is described in Sect.~4. The results are 
discussed in Sect.~5, and the concluding remarks are given in Sect.~6.

\section{Observations, data, and data reduction}

\subsection{New spectral line observations with APEX}

\subsubsection{Single-pointing observations}

A single-pointing position at $\alpha_{2000.0}=05^{\rm h}42^{\rm m}45\fs24$, and 
$\delta_{2000.0}=-01\degr16\arcmin14\farcs0$ (i.e. the \textit{Spitzer} 24~$\mu$m peak of SMM3) was observed with the 12-metre APEX 
telescope\footnote{{\tt http://www.apex-telescope.org/}} (\citealp{gusten2006}) 
in the frequency range $\sim218.2-222.2$ GHz. The observations were carried 
out on 20 August 2013, when the amount of precipitable water vapour (PWV) was 
measured to be 1.3~mm, which corresponds to a zenith atmospheric transmission 
of about 93\%. 

As a front end we used the APEX-1 receiver of the Swedish Heterodyne
Facility Instrument (SHeFI; \citealp{belitsky2007}; \citealp{vassilev2008a},b). 
The APEX-1 receiver operates in a single-sideband (SSB) mode using
sideband separation mixers, and it has a sideband rejection
ratio better than 10~dB. The backend was the RPG eXtended bandwidth
Fast Fourier Transfrom Spectrometer (XFFTS; see \citealp{klein2012}) with an 
instantaneous bandwidth of 2.5~GHz and 32\,768 spectral channels. The 
spectrometer consists of two units, which have a fixed overlap region of 
1.0~GHz. The resulting channel spacing, 76.3~kHz, corresponds to 
104~m~s$^{-1}$ at the central observed frequency of 220\,196.65~MHz. 
The beam size (Half-Power Beam Width or HPBW) at the observed frequency range 
is $\sim28\farcs1-28\farcs6$.

The observations were performed in the wobbler-switching
mode with a $100\arcsec$ azimuthal throw between two positions on sky 
(symmetric offsets), and a chopping rate of $R=0.5$ Hz. The total on-source 
integration time was 34 min. The telescope focus and pointing were optimised 
and checked at regular intervals on the planet Jupiter and the variable star 
R Leporis (Hind’s Crimson Star). The pointing was found to be accurate to 
$\sim3\arcsec$. The typical SSB system temperatures during the observations 
were in the range $T_{\rm sys}\sim130-140$ K. Calibration was made by means of 
the chopper-wheel technique, and the output intensity scale given by the system 
is the antenna temperature corrected for the atmospheric 
attenuation ($T_{\rm A}^{\star}$). The observed intensities were converted 
to the main-beam brightness temperature scale by 
$T_{\rm MB}=T_{\rm A}^{\star}/\eta_{\rm MB}$, where $\eta_{\rm MB}=0.75$ is the 
main-beam efficiency at the observed frequency range. The absolute calibration 
uncertainty is estimated to be about 10\%.

The spectra were reduced using the Continuum and Line Analysis 
Single-dish Software 90 ({\tt CLASS90}) program of the GILDAS software 
package\footnote{Grenoble Image and Line Data Analysis 
Software (GILDAS) is provided and actively developed by Institut de 
Radioastronomie Millim\'etrique (IRAM), and is available 
at {\tt http://www.iram.fr/IRAMFR/GILDAS}}. The individual spectra were
averaged, and the resulting spectra were Hanning-smoothed to a 
velocity resolution of 208~m~s$^{-1}$ to improve the 
signal-to-noise (S/N) ratio. Linear (first-order) baselines
were determined from the velocity ranges free of spectral line features, 
and then subtracted from the spectra. The resulting $1\sigma$ rms
noise levels at the smoothed velocity resolution were $\sim6.3-19$~mK on 
a $T_{\rm A}^{\star}$ scale, or $\sim8.4-25.3$~mK on a $T_{\rm MB}$ scale. 

The line identification from the observed frequency range was done by using 
{\tt Weeds}, which is an extension of {\tt CLASS} (\citealp{maret2011}), 
and the JPL\footnote{Jet Propulsion Laboratory (JPL) spectroscopic database 
(\citealp{pickett1998}); see {\tt http://spec.jpl.nasa.gov/}} 
and CDMS\footnote{Cologne Database for Molecular Spectroscopy 
(CDMS; \citealp{muller2005}); see {\tt http://www.astro.uni-koeln.de/cdms}} 
spectroscopic databases. The following spectral line transitions were 
detected: $^{13}$CO$(2-1)$, C$^{18}$O$(2-1)$, SO$(5_6-4_5)$, 
\textit{para}-H$_2$CO$(3_{0,\,3}-2_{0,\,2})$, 
\textit{para}-H$_2$CO$(3_{2,\,2}-2_{2,\,1})$, 
\textit{para}-H$_2$CO$(3_{2,\,1}-2_{2,\,0})$, and E$_1$-type CH$_3$OH$(4_2-3_1)$. 
Selected spectroscopic para\-meters of the detected species and transitions are 
given in Table~\ref{table:lines}. We note that the original purpose of these 
observations was to search for glycolaldehyde (HCOCH$_2$OH) line emission near 
220.2 GHz (see \citealp{jorgensen2012}; \citealp{coutens2015}). However, no positive detection of HCOCH$_2$OH lines was made. 

\subsubsection{Mapping observations}

The APEX telescope was also used to map SMM3 and its surroundings in the 
frequency range $\sim215.1-219.1$~GHz. The observations were done on 
15 November 2013, with the total telescope time of 2.9~hr. The target 
field, mapped using the total power on-the-fly mode, was 
$5\arcmin \times 3\farcm25$ ($0.61\times0.40$~pc$^2$) in size, and centred on 
the coordinates $\alpha_{2000.0}=05^{\rm h}42^{\rm m}47\fs071$, and 
$\delta_{2000.0}=-01\degr16\arcmin33\farcs70$. At the observed frequency range, the telescope HPBW is $\sim28\farcs5-29\arcsec$. The target area was scanned alternately in 
right ascension and declination, i.e. in zigzags to ensure minimal striping 
artefacts in the final data cubes. Both the angular separation between two 
successive dumps and the step size between the subscans was $9\farcs5$, i.e.
about one-third the HPBW. We note that to avoid beam smearing, 
the readout spacing should not exceed the value HPBW/3. The dump time was set 
to one second. The front end/backend system was composed of the APEX-1 
receiver, and the 2.5~GHz XFFTS with 32\,768 channels. The channel spacing, 
76.3~kHz, corresponds to 105~m~s$^{-1}$ at the central observed frequency 
of 217\,104.98~MHz.  

The focus and pointing measurements were carried out by making 
CO$(2-1)$ cross maps of the planet Jupiter and the M-type red supergiant 
$\alpha$ Orionis (Betelgeuse). The pointing was found to be consistent within 
$\sim3\arcsec$. The amount of PWV was $\sim0.6$~mm, which translates into 
a zenith transmission of about 96\%. The data were calibrated using the 
standard chopper-wheel method, and the typical SSB system temperatures 
during the observations were in the range $T_{\rm sys}\sim120-130$~K on a 
$T_{\rm A}^{\star}$ scale. The main-beam efficiency needed in the conversion 
to the main-beam brightness temperature scale is $\eta_{\rm MB}=0.75$. 
The absolute calibration uncertainty is about 10\%.

The {\tt CLASS90} program was used to reduce the spectra. 
The individual spectra were Hanning-smoothed to a velocity resolution of 
210~m~s$^{-1}$ to improve the S/N ratio of the data, and a third-order
polynomial was applied to correct the baseline in the spectra. 
The resulting $1\sigma$ rms noise level of the average smoothed spectra were 
about 90~mK on a $T_{\rm A}^{\star}$ scale. The visible spectral lines, 
identified by using {\tt Weeds}, were assigned to DCO$^+(3-2)$ and 
\textit{p}-H$_2$CO$(3_{0,\,3}-2_{0,\,2})$ (see Table~\ref{table:lines} for 
details). The latter line showed an additional velocity component at 
$v_{\rm LSR}\simeq1.5$~km~s$^{-1}$, while the systemic velocity of SMM3 
is about 8.5~km~s$^{-1}$. The main purpose of these mapping observations was 
to search for SiO$(5-4)$ emission at 217\,104.98~MHz, but no signatures of 
this shock tracer were detected. 

The spectral-line maps were produced using the Grenoble Graphic ({\tt GreG}) 
program of the GILDAS software package. The data were convolved with a 
Gaussian of 1/3 times the HPBW, and hence the effective angular 
resolutions of the final DCO$^+(3-2)$ and 
\textit{p}-H$_2$CO$(3_{0,\,3}-2_{0,\,2})$ data cubes are $30\farcs7$ and 
$30\farcs4$, respectively. The average $1\sigma$ rms noise level of the 
completed maps was $\sigma(T_{\rm MB})\sim100$~mK per 0.21~km~s$^{-1}$ channel. 

\subsection{Previous spectral line observations}

In the present work, we also employ the \textit{para}-NH$_3(1,\,1)$ and 
$(2,\,2)$ inversion line data obtained with the Effelsberg 100~m 
telescope\footnote{The 100~m telescope at Effelsberg/Germany is operated by 
the Max-Planck-Institut f\"ur Radioastronomie on behalf of the 
Max-Planck-Gesellschaft (MPG).} as described in Paper~II. The angular 
resolution (full-width at half maximum or FWHM) of these observations was 
$40\arcsec$. The original channel separation was 77~m~s$^{-1}$, but the spectra 
were smoothed to the velocity resolution of 154~m~s$^{-1}$. 
We note that the observed target position towards SMM3 was $\alpha_{2000.0}=05^{\rm h}42^{\rm m}44\fs4$, and $\delta_{2000.0}=-01\degr16\arcmin03\farcs0$, i.e. $\sim16\farcs7$ northwest of the new target position (Sect.~2.1.1).

In Paper~III, we presented the C$^{17}$O$(2-1)$, DCO$^+$(4-3), N$_2$H$^+(3-2)$, 
and N$_2$D$^+(3-2)$ observations carried out with APEX towards the aforementioned NH$_3$ target position. Here, we will employ these data as well. 
The HPBW of APEX at the frequencies of the above transitions is in the range 
$21\farcs7-27\farcs8$, and the smoothed velocity resolution is 260~m~s$^{-1}$ 
for N$_2$H$^+$ and DCO$^+$, and 320~m~s$^{-1}$ for C$^{17}$O and N$_2$D$^+$. For 
further details, we refer to Paper~III. Spectroscopic parameters of the 
species and transitions described in this subsection are also tabulated in 
Table~\ref{table:lines}.

%
%

\begin{table*}
\scriptsize
\caption{The observed molecular spectral lines and selected spectroscopic parameters.}
\label{table:lines}
\begin{tabular}{c c c c c c}
\hline\hline 
Transition & $\nu$ & $E_{\rm u}/k_{\rm B}$ & $\mu$ & $n_{\rm crit}$ & Rotational constants and  \\
      & [MHz] & [K] & [D] & [cm$^{-3}$] & Ray's parameter ($\kappa$) \\
\hline        
\textit{p}-NH$_3(J,\,K=1,\,1)$ & 23\,694.4955 & 23.26 & 1.4719 ($=\mu_C$) & $3.9\times10^3$\tablenotemark{a} & $A=B=298\,192.92$ MHz,     \\
                              &              &      &                  & & $C=186\,695.86$ MHz; \\
                               &              &      &                  & &  $\kappa=+1\Rightarrow$ oblate symmetric top\\
\textit{p}-NH$_3(J,\,K=2,\,2)$ & 23\,722.6333 & 64.45 & 1.4719 ($=\mu_C$) & $3.08\times10^3$\tablenotemark{a} & \ldots \\
DCO$^+(J=3-2)$ & 216\,112.5766\tablenotemark{b} & 20.74 & 3.888 ($=\mu_A$) & $2.0\times10^6$\tablenotemark{c} & $B=36\,019.76$ MHz; linear molecule\\
\textit{p}-H$_2$CO$(J_{K_a,\,K_c}=3_{0,\,3}-2_{0,\,2})$ & 218\,222.192 & 20.96 & 2.331 ($=\mu_A$) & $2.8\times10^6$\tablenotemark{c} & $A=281\,970.5$ MHz, $B=38\,833.98$ MHz,   \\
                                                  &              &     &                  & & $C=34\,004.24$ MHz; \\
                                                  &              &     &                  & & $\kappa=-0.961\Rightarrow$ prolate asymmetric top \\
CH$_3$OH-E$_1(J_{K_a,\,K_c}=4_{2,\,2}-3_{1,\,2})$ & 218\,440.050 & 45.46 & 0.899 ($=\mu_A$) & $4.7\times10^6$\tablenotemark{c} & $A=127\,523.4$ MHz, $B=24\,690.2$ MHz, \\
                           &              &       & $-1.44$ ($=\mu_B$) & & $C=23\,759.7$ MHz; \\
                           &              &       &                  & & $\kappa=-0.982\Rightarrow$ prolate asymmetric top\\
\textit{p}-H$_2$CO$(J_{K_a,\,K_c}=3_{2,\,2}-2_{2,\,1})$ & 218\,475.632 & 68.09 & 2.331 ($=\mu_A$) & $1.2\times10^6$\tablenotemark{c} & \ldots \\
\textit{p}-H$_2$CO$(J_{K_a,\,K_c}=3_{2,\,1}-2_{2,\,0})$ & 218\,760.066 & 68.11 & 2.331 ($=\mu_A$) & $2.6\times10^6$\tablenotemark{c} & \ldots \\
C$^{18}$O$(J=2-1)$ & 219\,560.3568 & 15.81 & 0.11079 ($=\mu_A$) & $2.0\times10^4$\tablenotemark{c} & $B=54\,891.42$ MHz; linear molecule\\
SO$(N_J=5_6-4_5)$ & 219\,949.442 & 34.98 & 1.55 ($=\mu_A$) & $2.4\times10^6$\tablenotemark{d} & $B=21\,523.02$ MHz; linear molecule\\
$^{13}$CO$(J=2-1)$ & 220\,398.7006\tablenotemark{e} & 15.87 & 0.11046 ($=\mu_A$) & $2.0\times10^4$\tablenotemark{c} & $B=55\,101.01$ MHz; linear molecule \\
C$^{17}$O$(J=2-1)$ & 224\,714.199\tablenotemark{f} & 16.18 & 0.11034 ($=\mu_A$) &$2.1\times10^4$\tablenotemark{c} & $B=56\,179.99$ MHz; linear molecule \\
N$_2$D$^+(J=3-2)$ & 231\,321.912\tablenotemark{g} & 22.20 & 3.40 ($=\mu_A$) & $1.9\times10^6$\tablenotemark{h} & $B=38\,554.71$ MHz; linear molecule \\
N$_2$H$^+(J=3-2)$ & 279\,511.832\tablenotemark{g} & 26.83 & 3.40 ($=\mu_A$) & $3.3\times10^6$\tablenotemark{h} & $B=46\,586.86$ MHz; linear molecule \\
DCO$^+(J=4-3)$ & 288\,143.855\tablenotemark{i} & 34.57 & 3.888 ($=\mu_A$) & $1.9\times10^7$\tablenotemark{c} & $B=36\,019.76$ MHz; linear molecule \\
\hline 
\end{tabular} 
\tablecomments{The spectroscopic data were compiled from the JPL database except in the cases of CH$_3$OH and $^{13}$CO where the data were taken from the CDMS. In columns (2)--(5) we list the rest frequency, upper-state energy divided by the Boltzmann constant, permanent electric dipole moment, and critical density at 10~K unless otherwise stated. In the last column, we give the rotational constants ($A,\,B,\,C$) and the Ray's asymmetry parameter, which is defined by $\kappa=(2B-A-C)/(A-C)$.}\tablenotetext{a}{From \citealp{maret2009}.}\tablenotetext{b}{Frequency of the strongest hyperfine component $F=4-3$ (JPL).}\tablenotetext{c}{To calculate $n_{\rm crit}$, we used the Einstein $A$ coefficients and collision rates ($C_{\rm ul}$) adopted from the Leiden Atomic and Molecular Database (LAMDA; \citealp{schoier2005}); {\tt http://home.strw.leidenuniv.nl/$\sim$moldata/}.}\tablenotetext{d}{A value of $C_{\rm ul}$ at 60~K from LAMDA was used (i.e. at the lowest temperature value reported in the database).}\tablenotetext{e}{Frequency of the strongest hyperfine component $F=5/2-3/2$ (CDMS).}\tablenotetext{f}{Frequency of the strongest hyperfine component $F=9/2-7/2$ (\citealp{ladd1998}).}\tablenotetext{g}{Frequency of the strongest hyperfine component $F_1,\,F=4,\,5-3,\,4$ (\citealp{pagani2009}; their Tables~4 and 10).}\tablenotetext{h}{To calculate $n_{\rm crit}$, we used the Einstein $A$ coefficients from Pagani et al. (2009), and the N$_2$H$^+$--H$_2$ collision rate from LAMDA.}\tablenotetext{i}{Frequency of the strongest hyperfine component $F=5-4$ (JPL).}
\end{table*}

\subsection{Submillimetre dust continuum data}

In the present study, we use our LABOCA 870~$\mu$m data 
first published in Paper~I. However, we have re-reduced 
the data using the Comprehensive Reduction Utility for SHARC-2 (Submillimetre 
High Angular Resolution Camera II) or CRUSH-2 (version 2.12-2) software package\footnote{{\tt http://www.submm.caltech.edu/$\sim$sharc/crush/index.htm}} (\citealp{kovacs2008}), as explained in more detail in the paper by Miettinen \& 
Offner (2013a). The resulting angular resolution was $19\farcs86$ (FWHM), 
and the $1\sigma$ rms noise level in the final map was 30~mJy~beam$^{-1}$. 
Measuring the flux density of SMM3 inside an aperture of radius equal to the 
effective beam FWHM, we obtained a value of $S_{\rm 870\,\mu m}=1.58\pm0.29$~Jy, 
where the uncertainty includes both the calibration uncertainty 
($\sim10\%$) and the map rms noise around the source (added in quadrature).

The SABOCA 350~$\mu$m data published in Paper~III are also used in this study. 
Those data were also reduced with CRUSH-2 (version 2.03-2). The obtained 
angular resolution was $10\farcs6$ (FWHM), and the $1\sigma$ rms noise was 
$\sim60$~mJy~beam$^{-1}$. Again, if the flux density is calculated using 
an aperture of radius $10\farcs6$, we obtain 
$S_{\rm 350\,\mu m}=4.23\pm1.30$~Jy, where the quoted error includes both 
the calibration uncertainty ($\sim30\%$) and the local rms noise. 
This value is about 1.3 times lower than the one reported in 
Paper~III ($5.4\pm1.6$~Jy, which was based on a clumpfind analysis above 
a $3\sigma$ emission threshold). The APEX dust continuum flux densities of SMM3 are tabulated in Table~\ref{table:photometry}.

\subsection{Far-infrared and millimetre data from the literature}

For the purpose of the present study, we use the far-infrared (FIR) 
flux densities from S13, and the 2.9~mm flux density of $S_{\rm 2.9\, mm}=115.4\pm3.9$~mJy from Tobin et al. (2015). Stutz et al. (2013) employed the \textit{Herschel}/Photodetector Array Camera \& Spectrometer (PACS; \citealp{pilbratt2010}; \citealp{poglitsch2010}) observations of SMM3 at 
70 and 160~$\mu$m. Moreover, they used the \textit{Herschel}/PACS 100~$\mu$m 
data from the HGBS. The aperture radii used for the photometry at the 
aforementioned three wavelengths were $9\farcs6$, $12\farcs8$, and $9\farcs6$, 
respectively, and the flux densities were found to be 
$S_{\rm 70\,\mu m}=3.29\pm0.16$~Jy, $S_{\rm 100\,\mu m}=10.91\pm2.79$~Jy, and 
$S_{\rm 160\,\mu m}=16.94\pm2.54$~Jy (see Table~4 in S13). 
We note that the \textit{Spitzer}/MIPS (the Multiband Imaging Photometer for 
\textit{Spitzer}; \citealp{rieke2004}) 70~$\mu$m flux density we determined 
in Paper~I, $3.6\pm0.4$~Jy, is consistent with the aforementioned 
\textit{Herschel}-based measurement (see Table~\ref{table:photometry} 
for the flux density comparison).

\begin{table*}
\caption{Mid-infrared to millimetre photometry of SMM3.}
\label{table:photometry}
\begin{tabular}{c c c c c c c c}
\hline\hline 
Reference & $S_{\rm 24\,\mu m}$ & $S_{\rm 70\,\mu m}$\tablenotemark{a} & $S_{\rm 100\,\mu m}$ & $S_{\rm 160\,\mu m}$ & $S_{\rm 350\,\mu m}$ & $S_{\rm 870\,\mu m}$ & $S_{\rm 2.9\,mm}$\\
      & [mJy] & [Jy] & [Jy] & [Jy] & [Jy] & [Jy] & [mJy] \\
\hline  
This work & \ldots & \ldots & \ldots & \ldots & $4.23\pm1.30$ & $1.58\pm0.29$ & \ldots\\
Paper~I & $5.0\pm0.2$ & $3.6\pm0.4$ & \ldots & \ldots & \ldots & $2.5\pm0.4$ & \ldots\\
Paper~III & \ldots & \ldots & \ldots & \ldots & $5.4\pm1.6$ & \ldots & \ldots\\
\citealp{stutz2013} & $4.74\pm0.3$ & $3.29\pm0.16$ & $10.91\pm2.79$ & $16.94\pm2.54$ & 3.63\tablenotemark{b} & 2.2/1.9\tablenotemark{c}& \ldots \\
\citealp{tobin2015} & \ldots & \ldots & \ldots & \ldots & \ldots & \ldots & $115.4\pm3.9$ \\
\hline 
\end{tabular} 
\tablecomments{See the reference studies and text herein for details on how the tabulated flux densities were measured.}\tablenotetext{a}{The 70~$\mu$m flux density from Paper~I was measured using the \textit{Spitzer}/MIPS data, while S13 used the \textit{Herschel}/PACS data.}\tablenotetext{b}{The authors adopted the SABOCA 350~$\mu$m peak surface brightness from Paper~III.}\tablenotetext{c}{The first value refers to a flux density measured in an aperture with radius equal to the beam FWHM ($19\arcsec$), while the latter one is otherwise the same but represents a background-subtracted value.}   
\end{table*}

\section{Observational results}

\subsection{Images of continuum emission}

\begin{figure*}
\begin{center}
\includegraphics[width=\textwidth]{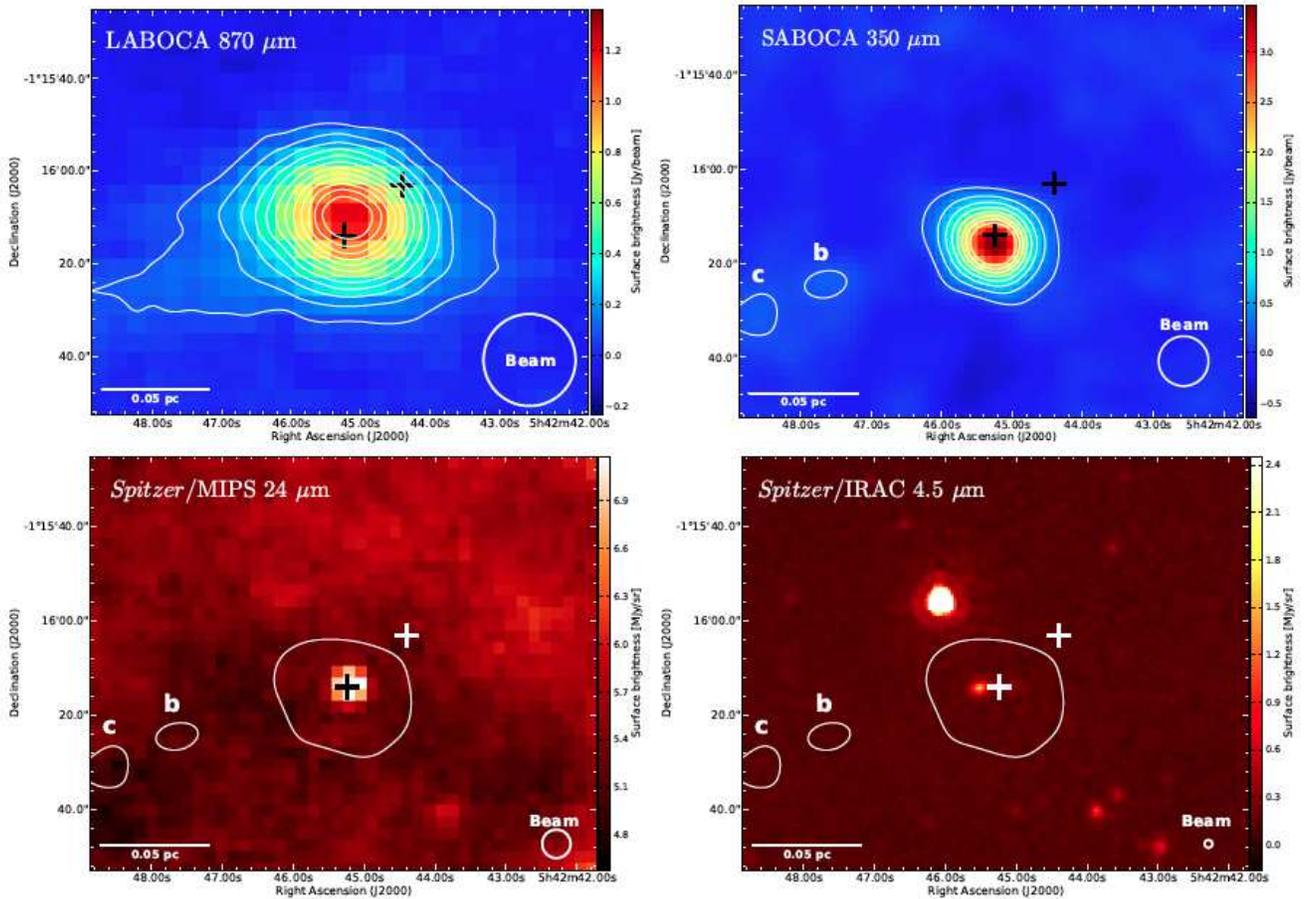}
\caption{Multiwavelength views of the SMM3 core. From top left to bottom right the panels show the LABOCA 870~$\mu$m, SABOCA 350~$\mu$m, \textit{Spitzer}/MIPS 24~$\mu$m, and \textit{Spitzer}/IRAC 4.5~$\mu$m images. The images are shown with linear scaling, and the colour bars indicate the surface-brightness scale in Jy~beam$^{-1}$ (APEX bolometers) or MJy~sr$^{-1}$ (\textit{Spitzer}).  The overlaid LABOCA contours in the top left panel start at $3\sigma$, and increase in steps of $3\sigma$, where $3\sigma=90$~mJy~beam$^{-1}$. The SABOCA contours also start at $3\sigma$, but increase in steps of $5\sigma$ ($1\sigma=60$~mJy~beam$^{-1}$). Both the \textit{Spitzer} images are overlaid with the $3\sigma$ SABOCA contours. The positions of our molecular line observations are marked by plus signs. The 350 $\mu$m condensations, SMM3b and 3c, are also indicated. In the bottom left corner of each panel, a scale bar indicating the 0.05~pc projected length is shown. In the bottom right corner of each panel, the circle shows the beam size (HPBW).}
\label{figure:images}
\end{center}
\end{figure*}

In Fig.~\ref{figure:images}, we show the SABOCA and LABOCA submm images of SMM3,
and \textit{Spitzer} 4.5~$\mu$m and 24~$\mu$m images of the same region. 
We note that the latter two were retrieved from a set of Enhanced Imaging 
Products (SEIP) from the \textit{Spitzer} Heritage Archive (SHA)\footnote{{\tt http://sha.ipac.caltech.edu/applications/Spitzer/SHA/}}, which include both the Infrared Array Camera (IRAC; \citealp{fazio2004}) and MIPS Super Mosaics.

The LABOCA 870~$\mu$m dust continuum emission is slightly extended to the 
east of the centrally concentrated part of the core. From this eastern part the 
SABOCA 350~$\mu$m image reveals the presence of two subcondensations, 
designated SMM3b and 3c (Paper~III). The \textit{Spitzer} 24~$\mu$m image 
clearly shows that the core harbours a central protostar, while the 4.5~$\mu$m 
feature slightly east of the 24~$\mu$m peak is probably related to shock 
emission. In particular, the 4.5~$\mu$m band is sensitive to shock-excited 
H$_2$ and CO spectral line features (e.g. \citealp{smith2005}; \citealp{ybarra2009}; \citealp{debuizer2010}). As indicated by the plus signs in 
Fig.~\ref{figure:images}, our previous line observations probed the outer edge 
of SMM3, i.e. the envelope region. In contrast, the present single pointing 
line observations were made towards the 24~$\mu$m peak position. This 
positional difference has to be taken into account when comparing the chemical 
properties derived from our spectral line data.

\subsection{Spectral line maps}

\begin{figure}[H]
\centering
\resizebox{\hsize}{!}{\includegraphics{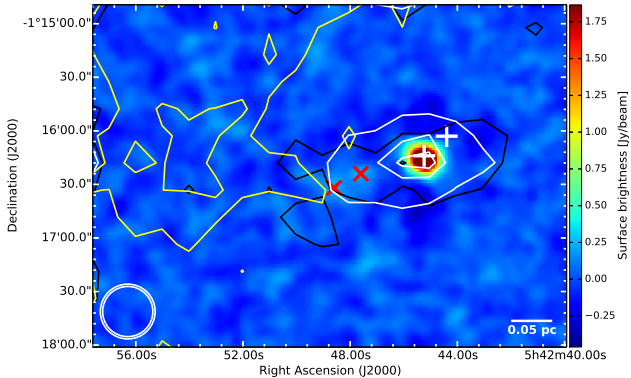}}
\caption{Spectral line emission maps overlaid on the SABOCA 350 
$\mu$m image. The DCO$^+(3-2)$ emission is shown with white contours plotted 
at $3\sigma$, $6\sigma$, and $7\sigma$ ($1\sigma=0.16$~K~km~s$^{-1}$). The 
black contours show the \textit{p}-H$_2$CO$(3_{0,\,3}-2_{0,\,2})$ emission, 
plotted at $3\sigma$ and $5\sigma$ ($1\sigma=0.1$~K~km~s$^{-1}$). The yellow 
contours, which show the low-velocity component of 
\textit{p}-H$_2$CO$(3_{0,\,3}-2_{0,\,2})$, are drawn at $3\sigma$ and $6\sigma$ 
($1\sigma=0.08$~K~km~s$^{-1}$). The white plus signs mark our new and previous 
single-pointing line observation positions. The red crosses indicate the 350~$\mu$m peak positions of SMM3b and 3c (Paper~III). The nested circles in 
the bottom left corner indicate the HPBW values of $28\farcs1$ and $30\farcs7$, 
i.e. the highest and coarsest resolutions of our new line observations 
(see Sect.~2). A scale bar indicating the 0.05~pc projected length is shown 
in the bottom right corner.}
\label{figure:linemaps}
\end{figure}

In Fig.~\ref{figure:linemaps}, we show the zeroth moment maps or integrated 
intensity maps of DCO$^+(3-2)$ and \textit{p}-H$_2$CO$(3_{0,\,3}-2_{0,\,2})$ 
plotted as contours on the SABOCA 350~$\mu$m image. The DCO$^+(3-2)$ map 
was constructed by integrating the line emission over the local standard of 
rest (LSR) velocity range of [7.4, 11.8]~km~s$^{-1}$. 
The \textit{p}-H$_2$CO$(3_{0,\,3}-2_{0,\,2})$ line showed two velocity 
components. The line emission associated with SMM3 was integrated over 
[7.5, 11]~km~s$^{-1}$, while that of the lower-velocity component 
($v_{\rm LSR}\simeq1.5$~km~s$^{-1}$) was integrated over [-0.27, 2.49]~km~s$^{-1}$. The aforementioned velocity intervals were determined from the average 
spectra. The final $1\sigma$ noise levels in the zeroth moment maps were in 
the range 0.08--0.16~K~km~s$^{-1}$ (on a $T_{\rm MB}$ scale).

With an offset of only 
$\Delta \alpha=-2\farcs6,\, \Delta \delta=3\farcs3$, the DCO$^+(3-2)$ emission 
maximum is well coincident with the 350~$\mu$m peak position of the 
core. The corresponding offset from our new line observation target position is 
$\Delta \alpha=-1\farcs7,\, \Delta \delta=5\farcs3$. 
Moreover, the emission is extended to the east (and slightly to the west), which
resembles the dust emission morphology traced by LABOCA. 

The \textit{p}-H$_2$CO$(3_{0,\,3}-2_{0,\,2})$ emission, shown by black contours 
in Fig.~\ref{figure:linemaps}, is even more elongated in the east-west 
direction than that of DCO$^+$. The emission peak is located inside the $7\sigma$ contour of DCO$^+(3-2)$ emission. We note that the 350 $\mu$m subcondensations SMM3b and 3c lie within the $3\sigma$ contour of both the line emissions. 

The low-velocity component of \textit{p}-H$_2$CO$(3_{0,\,3}-2_{0,\,2})$, with a 
radial velocity of about 1.5~km~s$^{-1}$, is concentrated on the east and 
northeast parts of the mapped region. This is exactly where the 
$^{13}$CO$(2-1)$ and C$^{18}$O$(2-1)$ line emissions at $\sim1.3$~km~s$^{-1}$ 
were found to be concentrated (\citealp{miettinen2012b}). As discussed by 
Miettinen (2012b), several other high-density tracer lines at a radial 
velocity of 1.3--1.9~km~s$^{-1}$ have been detected towards other cores in 
Orion B9 (Papers I--III). Hence, the detection of 
\textit{p}-H$_2$CO$(3_{0,\,3}-2_{0,\,2})$ emission at this low velocity comes as 
no surprise.

\subsection{Spectra and spectral line parameters}

\begin{figure}[H]
\centering
\resizebox{0.7\hsize}{!}{\includegraphics{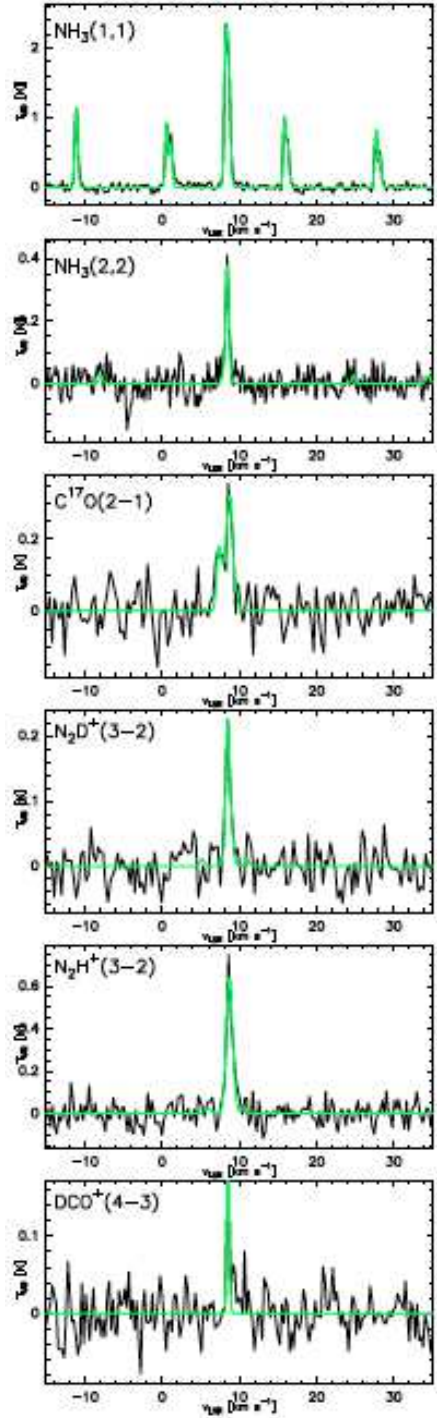}}
\caption{Hanning-smoothed spectra originally published in Papers~II and III. 
The hyperfine structure fits are shown with green lines. The velocity range 
shown in all panels was chosen so that the outer \textit{p}-NH$_3(1,\,1)$ 
satellite lines can be seen.}
\label{figure:spectra1}
\end{figure}

The previously observed spectra are shown in Fig.~\ref{figure:spectra1}. The 
target position of these measurements is shown by the northwestern plus sign 
in Figs.~\ref{figure:images} and \ref{figure:linemaps}. 

\begin{figure}[H]
\centering
\resizebox{0.6\hsize}{!}{\includegraphics{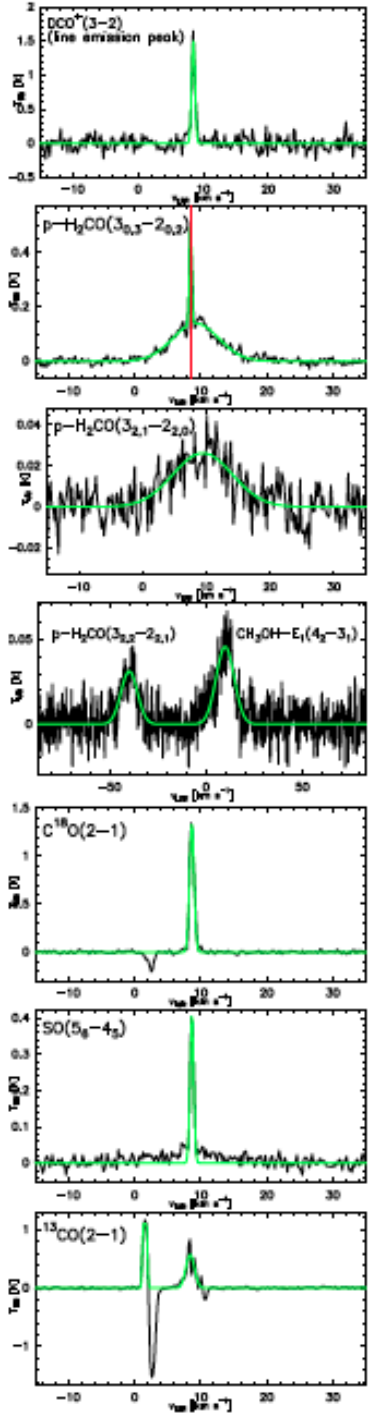}}
\caption{Hanning-smoothed spectra obtained with our new APEX observations. 
The DCO$^+(3-2)$ spectrum was extracted from the line emission peak.
The single-Gaussian fits are shown with green lines, while those overlaid on 
the DCO$^+$ and $^{13}$CO spectra show the hyperfine structure fits. The 
\textit{p}-H$_2$CO$(3_{0,\,3}-2_{0,\,2})$ and $^{13}$CO spectra show two velocity 
components. The red vertical line plotted on the former spectrum indicates 
the radial velocity of the \textit{p}-NH$_3(1,\,1)$ line. The velocity range 
is wider in the fourth panel from top to show the two nearby lines. The 
features with negative intensity in the C$^{18}$O and $^{13}$CO spectra are 
caused by emission in the observed OFF position.}
\label{figure:spectra2}
\end{figure}

The new spectra, observed towards the 24~$\mu$m peak of SMM3, are presented in 
Fig.~\ref{figure:spectra2}. The DCO$^+(3-2)$ spectrum shown in the top panel 
of Fig.~\ref{figure:spectra2} was extracted from the line emission peak, and, 
as mentioned above, that position is well coincident with the 24~$\mu$m and 
350~$\mu$m peaks (Fig.~\ref{figure:linemaps}). 
The \textit{p}-H$_2$CO$(3_{0,\,3}-2_{0,\,2})$ line shown in 
Fig.~\ref{figure:spectra2} can be decomposed into two components, 
namely a narrow line at the systemic velocity, and a much broader one with 
non-Gaussian line-wing emission. The narrow line is probably originating in 
the quiescent envelope around the protostar, while the broad component is 
probably tracing the dense ambient gas swept up by an outflow (e.g. 
\citealp{yildiz2013}). However, the mapped \textit{p}-H$_2$CO$(3_{0,\,3}-2_{0,\,2})$ data did not show evidence of line wings (and hence we could not separately image the blue and redshifted parts of the line emission). The other two formaldehyde lines ($3_{2,\,1}-2_{2,\,0}$ and $3_{2,\,2}-2_{2,\,1}$) and the CH$_3$OH line shown in Fig.~\ref{figure:spectra2} are also broad, and hence likely originate in the swept-up outflow gas. A hint of an outflow wing emission is also visible in the SO spectrum. 

Two velocity components are also seen in the $^{13}$CO spectrum, one at the systemic velocity, and the other at $\sim1.5$~km~s$^{-1}$, the velocity at which \textit{p}-H$_2$CO$(3_{0,\,3}-2_{0,\,2})$ emission was seen in the line maps. The C$^{18}$O and $^{13}$CO spectra exhibit absorption features next to the emission lines. These are caused by emission in the OFF beam positions when chopping between two positions on sky (wobbling secondary). This problem has been recognised in our previous papers on Orion B9, and is difficult to avoid when observing the abundant CO isotopologues. In fact, the detected $^{13}$CO line at the systemic velocity suffers so badly from the subtraction of the off-signal that the line shape and intensity are deformed. For example, the intensity of the C$^{18}$O line appears to be higher than that of the more abundant $^{13}$CO isotopologue. Hence, the $^{13}$CO data are not used in the present study.

The hyperfine structure of the ammonia lines were fitted using the 
{\tt CLASS90}'s methods NH3$(1,\,1)$ and NH3$(2,\,2)$. The former method 
could be used to derive the optical thickness of the main hyperfine group 
($\tau_{\rm m}$; see Sect.~4.2.1). The remaining lines 
shown in Fig.~\ref{figure:spectra1} are also split into hyperfine components, 
and hence were fitted using the {\tt CLASS90}'s hyperfine structure method. 

Of the newly observed lines, only DCO$^+(3-2)$ (cf.~\citealp{vandertak2009}) 
and $^{13}$CO$(2-1)$ (Cazzoli et al. 2004) exhibit hyperfine structure. 
In Fig.~\ref{figure:spectra2}, the fits to the $^{13}$CO lines are shown, but, 
as mentioned above, we do not study the lines further in the present paper. 
Single-Gaussian fits to the remaining lines were performed using {\tt CLASS90}. 
The obtained line parameters are listed in Table~\ref{table:lineparameters}. 
Columns~(2)--(5) in this table give the LSR velocity ($v_{\rm LSR}$), 
FWHM linewidth ($\Delta v$), peak intensity ($T_{\rm MB}$), and the 
integrated line intensity ($\int T_{\rm MB} {\rm d}v$). Besides the formal 
$1\sigma$ fitting errors, the errors in the last two quantities also include 
the calibration uncertainty (15\% for the Effelsberg/NH$_3$ data, and 10\% for 
our APEX data). We note that rather than using a Gaussian fit, the integrated 
intensity of the C$^{17}$O line was computed by integrating over the velocity 
range [5.87, 10.14]~km~s$^{-1}$ to take the non-Gaussian shape of the line 
into account.

\begin{table*}
\caption{Spectral line parameters.}
\footnotesize
\label{table:lineparameters}
\begin{tabular}{c c c c c c c c}
\hline\hline 
Transition & $v_{\rm LSR}$ & $\Delta v$ & $T_{\rm MB}$ & $\int T_{\rm MB} {\rm d}v$ & $\tau$ & $T_{\rm ex}$ & $T_{\rm rot}$ \\
     & [km~s$^{-1}$] & [km~s$^{-1}$] & [K] & [K~km~s$^{-1}$] & & [K] & [K]\\
\hline
\textit{p}-NH$_3(1,\,1)$ & $8.40\pm0.01$ & $0.40\pm0.01$ & $2.46\pm0.40$\tablenotemark{a} & $1.91\pm0.30$\tablenotemark{a} & $2.01\pm0.11\,(=\tau_{\rm m})$\tablenotemark{a} & $6.8\pm0.7$ & $10.6\pm0.5$ \\
\textit{p}-NH$_3(2,\,2)$ & $8.42\pm0.02$ & $0.45\pm0.08$ & $0.37\pm0.06$\tablenotemark{a} & $0.23\pm0.04$\tablenotemark{a} & $0.10\pm0.02\,(=\tau_0)$\tablenotemark{b} & $6.8\pm0.7$\tablenotemark{c} & \ldots\\
DCO$^+(3-2)$\tablenotemark{d} & $8.48\pm0.02$ & $0.60\pm0.04$ & $1.51\pm0.19$ & $0.98\pm0.11$ & $0.84\pm0.30\,(=\tau_0)$\tablenotemark{e} & $6.8\pm0.7$\tablenotemark{f} & \ldots \\
\textit{p}-H$_2$CO$(3_{0,\,3}-2_{0,\,2})$\tablenotemark{g} & $8.46\pm0.01$ & $0.42\pm0.02$ & $0.38\pm0.04$ & $0.17\pm0.02$ & $0.06\pm0.01\,(=\tau_0)$\tablenotemark{h} & $11.2\pm0.5$\tablenotemark{h} & \ldots \\
                                      & $9.26\pm0.08$ & $8.22\pm0.21$ & $0.14\pm0.02$ & $1.21\pm0.12$ & $\ll 1$\tablenotemark{i} & \ldots & $64\pm15$\tablenotemark{i}\\
CH$_3$OH-E$_1(4_{2,\,2}-3_{1,\,2})$ & $9.93\pm0.29$ & $10.98\pm0.80$ & $0.05\pm0.01$ & $0.54\pm0.06$ & $0.0009\pm0.0002\,(=\tau_0)$\tablenotemark{j} & $64\pm15$\tablenotemark{j} & \ldots \\
\textit{p}-H$_2$CO$(3_{2,\,2}-2_{2,\,1})$ & $8.99\pm0.47$ & $10.07\pm1.23$ & $0.03\pm0.01$ & $0.34\pm0.05$ & $\ll 1$\tablenotemark{i} & \ldots & $64\pm15$\tablenotemark{i}\\
\textit{p}-H$_2$CO$(3_{2,\,1}-2_{2,\,0})$ & $9.45\pm0.35$ & $10.92\pm0.83$ & $0.03\pm0.01$ & $0.31\pm0.04$ & $\ll 1$\tablenotemark{i} & \ldots & $64\pm15$\tablenotemark{i}\\
C$^{18}$O$(2-1)$ & $8.66\pm0.01$ & $0.82\pm0.01$ & $1.33\pm0.16$ & $1.17\pm0.12$ & $0.23\pm0.02\,(=\tau_0)$\tablenotemark{k} & $11.2\pm0.5$\tablenotemark{k} & \ldots\\
SO$(5_6-4_5)$ & $8.67\pm0.01$ & $0.68\pm0.02$ & $0.41\pm0.05$ & $0.29\pm0.03$ & $0.06\pm0.01\,(=\tau_0)$\tablenotemark{k} & $11.2\pm0.5$\tablenotemark{k} & \ldots\\
$^{13}$CO$(2-1)$\tablenotemark{g} & $8.38\pm0.15$ & $1.51\pm0.35$ & $0.57\pm0.18$ & $0.93\pm0.27$ & \ldots & \ldots & \ldots\\
                & $1.53\pm0.03$ & $0.43\pm0.03$ & $1.32\pm0.26$ & $0.89\pm0.15$ & \ldots & \ldots & \ldots \\
C$^{17}$O$(2-1)$ & $8.68\pm0.06$ & $0.59\pm0.11$ & $0.34\pm0.05$ & $0.54\pm0.07$ & $0.05\pm0.01\,(=\tau_0)$\tablenotemark{k} & $11.2\pm0.5$\tablenotemark{k} & \ldots\\
N$_2$D$^+(3-2)$ & $8.39\pm0.04$ & $0.53\pm0.11$ & $0.21\pm0.03$ & $0.17\pm0.03$ & $0.09\pm0.02\,(=\tau_0)$\tablenotemark{l} & $6.8\pm0.7$\tablenotemark{f} & \ldots\\
N$_2$H$^+(3-2)$ & $8.57\pm0.03$ & $0.85\pm0.09$ & $0.62\pm0.10$ & $0.67\pm0.08$ & $0.36\pm0.11\,(=\tau_0)$\tablenotemark{l} & $6.8\pm0.7$\tablenotemark{f} & \ldots\\
DCO$^+(4-3)$ & $8.54\pm0.03$ & $0.42\pm0.18$ & $0.20\pm0.02$ & $0.09\pm0.02$ & $0.11\pm0.03\,(=\tau_0)$\tablenotemark{e} & $6.8\pm0.7$\tablenotemark{f} & \ldots\\
\hline 
\end{tabular} 
\tablecomments{The parameters given in columns~(2)--(5) are described in Sect.~3.3, while those in the last three columns are the line optical thickness, excitation temperature, and rotational temperature (Sect.~4.2.1).}\tablenotetext{a}{These values refer to the main group of hyperfine components ($F_1=1-1$ and $F_1=2-2$ for the $(J,\,K)=(1,\,1)$ transition; $F_1=1-1$, $F_1=2-2$, and $F_1=3-3$ for the $(J,\,K)=(2,\,2)$ transition). The total NH$_3(1,\,1)$ line optical thickness is twice the main group value, i.e. $\tau_{\rm tot}=2\tau_{\rm m}$.}\tablenotetext{b}{Peak optical thickness of the strongest hyperfine component ($F=7/2-7/2$, $F_1=3-3$; weight $8/35$) calculated using $T_{\rm ex}[{\rm NH_3}(1,\,1)]$.}\tablenotetext{c}{Assumed to be that of the NH$_3(1,\,1)$ transition.}\tablenotetext{d}{The analysed beam-averaged spectrum was extracted from the line emission peak.}\tablenotetext{e}{The value of $\tau_0$ refers to the strongest hyperfine component, which is  $F=4-3$ for $J=3-2$ (relative intensity ${\rm R.I.}=3/7$), and $F=5-4$ for $J=4-3$ (${\rm R.I.}=11/27$).}\tablenotetext{f}{The value of $T_{\rm ex}$ was assumed to be that derived for NH$_3(1,\,1)$, and the value of $\tau_0$ was calculated based on this assumption.}\tablenotetext{g}{Two velocity components were detected. The $^{13}$CO lines are not analysed further in the present work (Sect.~3.3).}\tablenotetext{h}{The narrow line component at the systemic velocity was assumed to be thermalised at the kinetic temperature ($T_{\rm kin}$) derived from NH$_3$. The peak optical thickness was then calculated under the assumption that $T_{\rm ex}=T_{\rm kin}$.}\tablenotetext{i}{A rotational diagram method was used to derive $T_{\rm rot}$, under the assumption of optically thin emission.}\tablenotetext{j}{The value of $T_{\rm ex}$ was assumed to be equal to $T_{\rm rot}(p-{\rm H_2CO})$, and $\tau_0$ was estimated accordingly.}\tablenotetext{k}{The line was assumed to be thermalised at $T_{\rm kin}({\rm NH_3})$, and $\tau_0$ was calculated under this assumption. For C$^{17}$O, the relative intensity of the strongest hyperfine component $F=9/2-7/2$ is ${\rm R.I.}=1/3$.}\tablenotetext{l}{The value of $\tau_0$ refers to the strongest hyperfine component, i.e. $J_{F_1F}=3_{45}-2_{34}$ for both N$_2$H$^+$ and N$_2$D$^+$ (${\rm R.I.}=11/63$).}
\end{table*}

\section{Analysis and results}

\subsection{Spectral energy distribution of SMM3 -- modified blackbody fitting}

\begin{figure}[H]
\centering
\resizebox{\hsize}{!}{\includegraphics{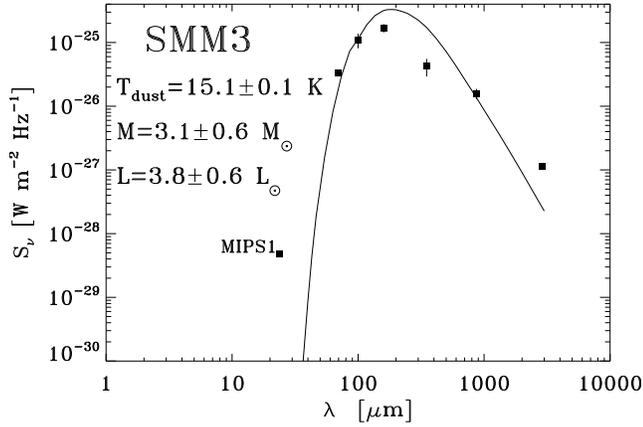}}
\caption{Spectral energy distribution of SMM3. The square symbols with vertical error bars represent the measured flux densities (\textit{Herschel}/PACS, SABOCA, LABOCA, and CARMA). A modified blackbody fit to the data points is shown 
by a solid black line. The \textit{Spitzer} 24~$\mu$m data point from S13 is also indicated (MIPS1), but not used in the fit.}
\label{figure:SED}
\end{figure}

The SED of SMM3, constructed using the \textit{Herschel}/ PACS 70, 100, 
and 160~$\mu$m, SABOCA 350~$\mu$m, LABOCA 870~$\mu$m, and CARMA 2.9~mm 
flux densities (see Sects.~2.3 and 2.4, and Table~\ref{table:photometry}), 
is show in Fig.~\ref{figure:SED}. The \textit{Spitzer} 24~$\mu$m data point, 
which represents a flux density of 
$4.74\pm0.3$~mJy from S13, is also shown in the figure, 
but it was excluded from the fit (see below). We note that the S13 24~$\mu$m 
flux density is close to a value of $5.0\pm0.2$~mJy we determined in Paper~I ($13\arcsec$ aperture; see Table~\ref{table:photometry}). The 24~$\mu$m emission 
originates in a warmer dust component closer to the accreting central 
protostar, while the longer wavelength data ($\lambda \geq70$~$\mu$m) 
are presumable tracing the colder envelope.

The solid line in Fig.~\ref{figure:SED} represents a single-temperature MBB 
function fitted to the aforementioned data points. 
The fit was accomplished with optimisation ($\chi^2$ minimisation) by 
simulated annealing (Kirkpatrick et al. 1983), which, although more time-consuming, can work better in finding the best fit solution than the most commonly-used standard (non-linear) least-squares fitting method that can be sensitive to the chosen initial values (see also \citealp{bertsimas1993}; \citealp{ireland2007}). The original version of the fitting algorithm was written by J.~Steinacker (M.~Hennemann, priv.~comm.). 
It was assumed that the thermal dust emission is optically thin ($\tau \ll 1$).
We note that this assumption is probably good for the wavelengths longward 
of 70~$\mu$m, but it gets worse at shorter wavelengths. This, together with the fact that 24~$\mu$m emission originates in a warmer dust component closer to 
the accreting central protostar than the longer wavelength emission ($\lambda \geq70$ $\mu$m) arising from the colder envelope, is the reason 
why we excluded the 24~$\mu$m flux density from the fit 
(e.g. \citealp{ragan2012}). The model fit takes into account the 
wavelength-dependence of the dust opacity ($\kappa_{\lambda}$). As the dust 
model, we employed the widely used Ossenkopf \& Henning (1994, hereafter OH94) model 
describing graphite-silicate dust grains that have coagulated and accreted 
thin ice mantles over a period of $10^5$~yr at a gas density of $10^5$~cm$^{-3}$. For the total dust-to-gas mass ratio we adopted a value of 
$\delta_{\rm dg}\equiv M_{\rm dust}/M_{\rm gas}=1/141$. This mass ratio is based on the assumption that 
the core's chemical composition is similar to the solar mixture, i.e. the mass fractions for hydrogen, helium, 
and heavier elements were assumed to be $X=0.71$, $Y=0.27$, and $Z=0.02$, 
respectively\footnote{In this case, the ratio between the total mass 
(H+He+metals) to hydrogen mass is $1/X\simeq1.41$.}.  

As can be seen in Fig.~\ref{figure:SED}, the PACS data are reasonably well fitted although the 
160~$\mu$m flux density is slightly overestimated. The SABOCA data point is 
not well fitted, which could be partly casused by the spatial filtering owing 
to the sky-noise removal. Hence, a ground-based bolometer flux density can 
appear lower than what would be expected from the \textit{Herschel} data. 
On the other hand, our LABOCA data point is well matched with the MBB fit. Finally, we note that the CARMA 2.9~mm flux density, which is based on the highest angular resolution data used here, is underestimated by the MBB curve. Radio continuum observations would be needed to quantify the amount of free-free contribution at 2.9~mm (cf.~\citealp{wardthompson2011}).  

The dust temperature, envelope mass, and luminosity obtained from the SED fit 
are $T_{\rm dust}=15.1\pm0.1$~K, $M_{\rm env}=3.1\pm0.6$~M$_{\sun}$, and 
$L=3.8\pm0.6$~L$_{\sun}$. However, we emphasise that these values should be taken with some caution because clearly the fit shown in Fig.~\ref{figure:SED} is not perfect. In principle, while the 24~$\mu$m emission is expected to trace a warmer dust component than those probed by $\lambda_{\rm obs} \geq70$~$\mu$m observations (e.g. \citealp{ragan2012}), it is possible that our poor single-$T_{\rm dust}$ fit reflects the presence of more than one cold dust components in the protostar's envelope, and would hence require a multi-$T_{\rm dust}$ fit. However, following S13, and to allow an easier comparison with their results, we opt to use a simplified single-$T_{\rm dust}$ MBB in the present study.

We note that $M_{\rm env}\propto (\kappa_{\lambda}\delta_{\rm dg})^{-1}$, and hence the choice of the dust model (effectively $\kappa_{\lambda}$) and $\delta_{\rm dg}$ mostly affect the envelope mass among the SED parameters derived here (by a factor of two or more; OH94). The adopted dust model can also (slightly) influence the derived values of $T_{\rm dust}$ and $L$ because of the varying dust emissivity index ($\beta$) among the different OH94 models ($\kappa_{\lambda}\propto \lambda^{-\beta}$). The submm luminosity, $L_{\rm submm}$, computed by numerically integrating the fitted SED curve longward of 350~$\mu$m, is about 0.23~L$_{\sun}$, i.e. about $6\%$ of the total luminosity. For Class 0 protostellar cores, the $L_{\rm submm}/L$ ratio is defined to be 
$>5\times10^{-3}$, which reflects the condition that the envelope mass exceeds 
that of the central protostar, i.e. $M_{\rm env}\gg M_{\star}$ (\citealp{andre1993}, 2000). With a $L_{\rm submm}/L$ ratio of about one order of magnitude higher 
than the definition limit, SMM3 is clearly in the Class 0 regime. 

Our $T_{\rm dust}$ value is by a factor of 1.4 lower than that obtained by S13 through their MBB analysis, while the values of $M_{\rm env}$ and $L$ we derived are higher by factors of about 9.4 and 1.8, 
respectively (see Sect.~1). We note that similarly to the present work, 
S13 fitted the data at $\lambda \geq 70$ $\mu$m, but they adopted a slightly different OH94 dust model (coagulation at a density of 
$10^6$~cm$^{-3}$ rather than at $10^5$~cm$^{-3}$ as here), 
and a slightly higher gas-to-dust ratio than we ($1.36\times110=149.6$, which 
is $6\%$ higher than our value of 141). Hence, we attribute the aforementioned discrepancies to the different SABOCA and LABOCA flux density values used in the analysis (e.g. S13 used the peak surface brightness from our SABOCA map, and their fit underestimated the LABOCA flux density), and to the fact that we have here used the new CARMA 2.9~mm data from Tobin et al. (2015) as well. 

Given that Class~0 objects have, by definition, $M_{\rm env}\gg M_{\star}$, an envelope mass of $\sim3$~M$_{\sun}$ derived here might be closer to the true value than a value of $\sim0.3$~M$_{\sun}$ derived by S13. Also, as was already mentioned in Sect.~1, SMM3 was found to be a very bright 2.9~mm-emitter by Tobin et al. (2015), and hence they derived a high mass of $7.0\pm0.7$~M$_{\sun}$ under the assumption that $T_{\rm dust}=20$~K and $\delta_{\rm dg}=1/100$ (their mass is $2.3\pm0.5$ times higher than the present estimate, but a direct comparison with a single-flux density analysis is not feasible). In the context of stellar evolution, if the core star formation efficiency is $\sim30$\% (e.g. \citealp{alves2007}), and the central SMM3 protostar has $M_{\star} \ll M_{\rm env}$, this source could evolve into a near solar-mass star if $M_{\rm env}\sim3$~M$_{\sun}$ as estimated here, while an envelope mass of $\sim0.3$~M$_{\sun}$ would only be sufficient to form a very low-mass single star (near the substellar--stellar limit of $\sim0.1$~M$_{\sun}$). Moreover, the dust temperature we have derived here is closer to the gas kinetic temperature in SMM3 (the ratio between the two is $1.35\pm0.06$; see Sect.~4.2.1) than the value $T_{\rm dust}=21.4\pm0.4$~K from S13. In a high-density protostellar 
envelope, the gas temperature is indeed expected to be similar to $T_{\rm dust}$ (e.g. the dust--gas coupling occurs at $\sim10^5$~cm$^{-3}$ in the 
\citealp{hollenbach1989} prescription). Finally, the physical implication of the higher luminosity we have derived here -- $1.8\pm0.3$ times the S13 value -- is that 
the mass accretion rate of the SMM3 protostar is higher by a similar factor.

\begin{figure}[H]
\centering
\resizebox{\hsize}{!}{\includegraphics{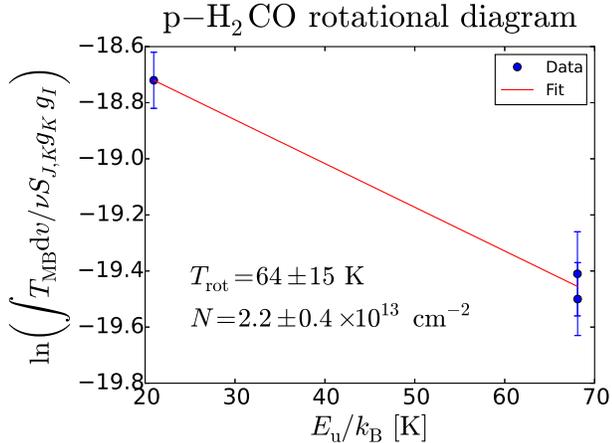}}
\caption{Rotational diagram for \textit{p}-H$_2$CO. The left-hand side of 
Eq.~(\ref{eq:rot}) is plotted as a function of the energy of the upper level.
The red solid line shows a least-squares fit to the observed data. The resulting values of 
$T_{\rm rot}$ and $N$ are indicated.}
\label{figure:rot}
\end{figure}

\subsection{Analysis of the spectral line data}

\subsubsection{Line optical thicknesses, and the excitation, rotational, and 
kinetic temperatures}

The optical thickness of the main \textit{p}-NH$_3(1,\,1)$ hyperfine group, $\tau_{\rm m}$, could be derived by fitting the hyperfine structure of the line. The main hyperfine group ($\Delta F=0$) has a relative strength of half the total value, and hence the total optical thickness of 
\textit{p}-NH$_3(1,\,1)$ is given by $\tau_{\rm tot}=2\tau_{\rm m}$ 
($=2\times(2.01\pm0.11)$; see \citealp{mangum1992}; Appendix~A1 
therein). The strongest hyperfine component has a relative strength of 
$7/30$, which corresponds to a peak optical thickness of $\tau_0\simeq0.94$. 
The excitation temperature of the line, $T_{\rm ex}$, was calculated from 
the antenna equation ($T_{\rm MB}\propto (1-e^{-\tau})$; see e.g Eq.~(1) 
in Paper~I), assuming that the background tempe\-rature is equal to that of the cosmic microwave
background radiation, i.e. $T_{\rm bg}\equiv T_{\rm CMB}=2.725$~K (\citealp{fixsen2009}). 
The obtained value, $T_{\rm ex}=6.8\pm0.7$~K\footnote{We note that 
in Paper~II we determined a value of $T_{\rm ex}({\rm NH_3})=6.1\pm0.5$~K from 
a unsmoothed \textit{p}-NH$_3(1,\,1)$ spectrum, while the present value 
was derived from a smoothed spectrum.}, was also adopted for the \textit{p}-NH$_3(2,\,2)$ line 
because its hyperfine satellites were not detected. Using this assumption and 
the antenna equation, the peak \textit{p}-NH$_3(2,\,2)$ optical thickness was 
determined to be $0.1\pm0.02$. To calculate $\tau_{\rm tot}$, this value should 
be scaled by the relative strength of the strongest hyperfine component which 
is $8/35$. The value $T_{\rm ex}=6.8\pm0.7$~K was also adopted for the 
N$_2$H$^+$, N$_2$D$^+$, and DCO$^+$ lines, although we note that they might 
originate in a denser gas than the observed ammonia lines.
Another caveat is that the $J=3-2$ line of DCO$^+$ was extracted from a 
position different from the ammonia target position, but, within the errors, 
the aforementioned $T_{\rm ex}$ value is expected to be a reasonable choice (e.g. 
\citealp{anderson1999}). The values of $\tau_0$ were then derived as in the case 
of the $(2,\,2)$ transition of ammonia (see Col.~(6) in 
Table~\ref{table:lineparameters}).

Using the $\tau_{\rm m}[p-{\rm NH_3(1,\,1)}]$ value and the intensity ratio 
between the $(2,\,2)$ and $(1,\,1)$ lines of \textit{p}-NH$_3$, we derived the 
rotational temperature of ammonia ($T_{\rm rot}$; see Eq.~(4) in 
\citealp{ho1979}). This calculation assumed that the $T_{\rm ex}$ values, and also the linewidths, 
are equal between the two inversion lines. The latter assumption is justified by the 
observed FWHM linewidths. The derived value of $T_{\rm rot}$, $10.6\pm0.5$~K, 
was converted into an estimate of the gas kinetic temperature using the 
$T_{\rm kin}-T_{\rm rot}$ relationship from Tafalla et al. (2004; their 
Appendix~B), which is valid in the low-temperature regime of 
$T_{\rm kin}\in [5,\,20]$~K. The value we derived, $T_{\rm kin}=11.2\pm0.5$ K\footnote{The quoted value of $T_{\rm kin}$ differs slightly from the one derived 
in Paper II ($11.3\pm0.8$~K) because of the smoothed ammonia spectra employed 
in the analysis in the present work.}, was adopted as $T_{\rm ex}$ for the observed 
CO isotopologue transitions, SO, and the narrow \textit{p}-H$_2$CO line. 
The choice of $T_{\rm ex}=T_{\rm kin}$ means that the level populations 
are assumed to be thermalised, and this is often done in the case of 
C$^{18}$O (e.g. \citealp{hacar2011}), while in the cases of SO and H$_2$CO it 
should be taken as a rough estimate only.

The three broad \textit{p}-H$_2$CO lines we detected allowed us to construct 
a rotational diagram for \textit{p}-H$_2$CO. The rotational diagram technique 
is well established, and details of the method can be found in a number of 
papers (e.g. \citealp{linke1979}; \citealp{turner1991}; 
\citealp{goldsmith1999}; Anderson et al. 1999; Green et al. 2013). When the line emission is assumed to 
be optically thin, the integrated intensity of the line is related to 
$T_{\rm rot}$ and the total column density of the species, $N$, according to the 
equation

\begin{equation}
\label{eq:rot}
\ln \left[\frac{\int T_{\rm MB}{\rm d}v}{\nu Sg_Kg_I}\right]=\ln \left(\frac{2\pi^2\mu^2}{3k_{\rm B}\epsilon_0}\frac{N}{Z_{\rm rot}}\right)-\frac{1}{T_{\rm rot}}\frac{E_{\rm u}}{k_{\rm B}}\,,
\end{equation}
where $S$ is the line strength, $g_K$ is the $K$-level 
degeneracy, $g_I$ is the reduced nuclear spin degeneracy, $\epsilon_0$ is the 
vacuum permitti\-vity, and $Z_{\rm rot}$ is the rotational partition function. 
The values of $S$ were adopted from the Splatalogue 
database\footnote{{\tt http://www.cv.nrao.edu/php/splat/}}. Because H$_2$CO is 
an asymmetric top molecule, there is no $K$-level degeneracy, and hence 
$g_K=1$. For the \textit{para} form of H$_2$CO ($K_a$ is even), the value 
of $g_I$ is $1/4$ (\citealp{turner1991}). The H$_2$CO molecule belongs to 
a $C_{2v}$ symmetry group (two vertical mirror planes), and its partition function 
at the high-temperature limit ($hA/k_{\rm B}T_{\rm ex} \ll 1$, where $h$ is the Planck 
constant) can be approximated as (\citealp{turner1991})

\begin{equation}
\label{eq:part}
Z_{\rm rot}(T_{\rm rot})\simeq\frac{1}{2}\sqrt{\frac{\pi(k_{\rm B}T_{\rm rot})^3}{h^3ABC}}\,.
\end{equation}
The derived rotational diagram, i.e. the left-hand side of Eq.~(\ref{eq:rot}) 
plotted as a function of $E_{\rm u}/k_{\rm B}$, is shown 
in Fig.~\ref{figure:rot}. The red solid line represents a least-squares fit 
to the three data points. The fit provides a value of $T_{\rm rot}$ as the 
reciprocal of the slope of the line, and $N$ can be calculated from the 
$y$-intercept. We note that two of the detected \textit{p}-H$_2$CO transitions have almost the same upper-state energy, i.e. they lie very close to each other in the direction of the $x$-axis in Fig.~\ref{figure:rot}, which makes the fitting results rather poorly constrained. We also note that the 
\textit{ortho}-H$_2$CO$(2_{1,\,1}-1_{1,\,1})$ line detected by Kang et al. 
(2015) refers to the narrow-line component ($\Delta v=0.45$~km~s$^{-1}$), and hence cannot be employed in our rotational diagram for the broad-line component. The value of $T_{\rm rot}$ we derived is $64\pm15$~K, which in the case of local thermodynamic equilibrium (LTE) is equal to $T_{\rm kin}$. Owing to the common formation route for formaldehyde and methanol (Sect.~5.2.3), the aforementioned $T_{\rm rot}$ value was adopted as $T_{\rm ex}$ for the detected CH$_3$OH line (which then appears to be optically thin). The molecular column density calculations are described in the next subsection. 

\subsubsection{Molecular column densities and fractional abundances}

As described above, the beam-averaged column density of \textit{p}-H$_2$CO 
for the broad component was derived using the rotational diagram method. 
The column densities of the species other than NH$_3$ (see below) were 
calculated by using the standard LTE formulation

\begin{equation}
\label{eq:N}
N=\frac{3h\epsilon_0}{2\pi^2}\frac{1}{\mu^2S}\frac{Z_{\rm rot}(T_{\rm ex})}{g_Kg_I}e^{E_u/k_{\rm B}T_{\rm ex}}F(T_{\rm ex})\int \tau(v){\rm d}v \, , 
\end{equation}
where $F(T_{\rm ex})\equiv \left(e^{h\nu/k_{\rm B}T_{\rm ex}}-1\right)^{-1}$. 
Here, the electric dipole moment matrix element is defined as 
$\left|\mu_{\rm ul} \right|\equiv \mu^2S/g_{\rm u}$, where $g_{\rm u}\equiv g_J=2J+1$ 
is the rotational degeneracy of the upper state (\citealp{townes1975}).     
The values of the product $\mu^2S$ were taken from the Splatalogue database, but
we note that for linear molecules $S$ is simply equal to the rotational quantum 
number of the upper state, i.e. $S=J$ (the SO molecule, which possesses a 
$^3\Sigma$ (electronic spin is 1) electronic ground state, is an exception; 
\citealp{tiemann1974}). For linear molecules, $g_K=g_I=1$ for all levels, while 
for the E-type CH$_3$OH, $g_K=2$ and $g_I=1$ (\citealp{turner1991}).

The partition function of the linear molecules was approximated as 

\begin{equation}
\label{eq:Z1}
Z_{\rm rot}(T_{\rm ex}) \simeq \frac{k_{\rm B}T_{\rm ex}}{hB}+\frac{1}{3}\,.
\end{equation}
Equation~(\ref{eq:Z1}) is appropriate for heteropolar molecules at a 
high-temperature limit of $hB/k_{\rm B}T_{\rm ex} \ll 1$. For SO, however, the 
rotational levels with $N\geq1$ are split into three sublevels (triplet of 
$N=J-1$, $N=J$, and $N=J+1$). To calculate the partition function of SO, 
we used the approximation formulae from Kontinen et al. (2000; Appendix~A 
therein). For CH$_3$OH, which has an internal rotor, the partition function 
is otherwise similar to that in Eq.~(\ref{eq:part}) but with a numerical 
factor of 2 instead of 1/2 (\citealp{turner1991}).

When the spectral line has a Gaussian profile, the last integral term in 
Eq.~(\ref{eq:N}) can be expressed as a function of the FWHM linewidth and peak 
optical thickness of the line as

\begin{equation}
\label{eq:tau}
\int \tau(v){\rm d}v=\frac{\sqrt{\pi}}{2\sqrt{\ln 2}}\Delta v \tau_0 \simeq1.064\Delta v \tau_0 \,.
\end{equation}
We note that for the lines with hyperfine structure the total optical 
thickness is the sum of peak optical thicknesses of the different components. 
Moreover, if the line emission is optically thin ($\tau \ll 1$), 
$T_{\rm MB}\propto \tau$, and $N$ can be computed from the integrated 
line intensity. The values of $\tau$ listed in Col.~(6) in 
Table~\ref{table:lineparameters} were used to decide whether the assumption 
of optically thin emission is valid (in which case the column density was 
calculated from the integrated intensity).

To derive the total column density of NH$_3$, we first calculated that in the 
$(1,\,1)$ state, which, by taking into account both parity states of the level,
is given by (e.g. \citealp{harju1993})

\begin{equation}
\label{eq:ammonia1}
N({\rm NH_3})_{(1,\,1)}=N_++N_-=N_+(1+e^{h\nu_{(1,\,1)}/k_{\rm B}T_{\rm ex}})\,.
\end{equation}
The latter equality follows from the Boltzmann population distribution, and 
the fact that the two levels have the same statistical weights ($J$ and $K$ do 
not change in the inversion transition). Because $N_+$ represents the column 
density in the upper state, its value was calculated from a formula that can 
be derived by substituting Eq.~(\ref{eq:tau}) into Eq.~(\ref{eq:N}), and
dividing by the term $Z_{\rm rot}/(g_Kg_I)e^{E_u/k_{\rm B}T_{\rm ex}}$. 
The value of $S$ for a $(J,\,K)\rightarrow(J,\,K)$ transition is 
$S=K^2/[J(J+1)]$. Finally, making 
the assumption that at the low temperature of SMM3 only the four lowest 
metastable ($J=K$) levels are populated, the value of $N({\rm NH_3})_{(1,\,1)}$ 
was scaled by the partition function ratio $Z_{\rm rot}/Z_{\rm rot}(1,\,1)$ to 
derive the total (\textit{ortho}+\textit{para}) NH$_3$ column density as 

\begin{eqnarray}
N({\rm NH_3}) &=& N({\rm NH_3})_{(0,\,0)}+N({\rm NH_3})_{(1,\,1)}\nonumber \\
& & +N({\rm NH_3})_{(2,\,2)}+ \,N({\rm NH_3})_{(3,\,3)}\nonumber \\
              &=& N({\rm NH_3})_{(1,\,1)}\times \nonumber \\
& & \left(\frac{1}{3}e^{\frac{23.4}{T_{\rm rot}}}+1+\frac{5}{3}e^{-\frac{41.5}{T_{\rm rot}}}+\frac{14}{3}e^{-\frac{101.2}{T_{\rm rot}}} \right)\,.
\end{eqnarray}

The column density analysis presented here assumes that 
the line emission fills the telescope beam, 
i.e. that the beam filling factor is unity. As can be seen in 
Fig.~\ref{figure:linemaps}, the DCO$^+(3-2)$ and 
\textit{p}-H$_2$CO$(3_{0,\,3}-2_{0,\,2})$ emissions are somewhat extended with 
respect to the 350~$\mu$m-emitting core whose size is comparable 
to the beam size of most of our line observations. Moreover, 
the detected N-bearing species are often found
to show spatial distributions comparable to the dust emission of dense cores 
(e.g. \citealp{caselli2002a}; \citealp{lai2003}; \citealp{daniel2013}). It is still 
possible, however, that the assumption of unity filling factor is not correct.
The gas within the beam area can be structured in a clumpy fashion, in which 
case the true filling factor is $<1$. The derived beam-averaged column density
is then only a lower limit to the source-averaged value.

The fractional abundances of the molecules were calculated by dividing the 
molecular column density by the H$_2$ column density, $x=N/N({\rm H_2})$. 
To be directly comparable to the molecular line data, the $N({\rm H_2})$ values
were derived from the LABOCA data smoothed to the resolution of 
the line observations (cf.~Eq.~(3) in Paper~I). For this calculation, we 
adopted the dust temperature derived from the SED fit ($T_{\rm dust}=15.1 \pm 0.1$~K), except for the broad component of \textit{p}-H$_2$CO and CH$_3$OH for which $T_{\rm dust}$ was assumed to be $64\pm15$~K ($=T_{\rm rot}(p-{\rm H_2CO})$). The mean molecular weight per H$_2$ molecule we used was $\mu_{\rm H_2}=2.82$, and the 
dust opacity per unit dust mass at 870~$\mu$m was set to
$\kappa_{\rm 870\,\mu m}=1.38$~cm$^2$~g$^{-1}$ to be consistent with the OH94 dust 
model described earlier. The beam-averaged column densities and abundances with 
respect to H$_2$ are listed in Table~\ref{table:chemistry}. 

\subsubsection{Deuterium fractionation and CO depletion}

The degree of deuterium fractionation in N$_2$H$^+$ was calculated by dividing 
the column density of N$_2$D$^+$ by that of N$_2$H$^+$. The obtained value, 
$14\%\pm6\%$, is about $40\%$ of the value derived in Paper~III (i.e. 
$0.338\pm0.09$ based on a non-LTE analysis). 

To estimate the amount by which the CO molecules are depleted in SMM3,
we calculated the CO depletion factors following the analysis presented in 
Paper~III with the following modifications. Recently, Ripple et al. (2013) 
analysed the CO abundance variation across the Orion giant molecular clouds.
In particular, they derived the $^{13}$CO fractional abundances, and found that
in the self-shielded interiors ($3<A_{\rm V}<10$ mag) of Orion B, the value 
of $x({\rm ^{13}CO})$ is $\simeq3.4\times10^{-6}$. On the other hand, towards 
NGC~2024 in Orion~B the average [$^{12}$C]/[$^{13}$C] ratio is measured to be 
about 68 (\citealp{savage2002}; \citealp{milam2005}). These two values translate 
into a canonical (or undepleted) CO abundance of $\simeq2.3\times10^{-4}$. 
We note that this is 2.3 times higher than the classic value $10^{-4}$, but 
fully consistent with the best-fitting CO abundance of $2.7_{-1.2}^{+6.4}\times10^{-4}$ 
found by Lacy et al. (1994) towards NGC 2024. Because we derived the C$^{18}$O 
and C$^{17}$O abundances towards the core centre and the envelope, respectively, 
the canonical abundances of these two species had to be estimated. We assumed 
that the [$^{16}$O]/[$^{18}$O] ratio is equal to the average local interstellar 
medium value of 557 (\citealp{wilson1999}), and that the [$^{18}$O]/[$^{17}$O] 
ratio is that derived by Wouterloot et al. (2008) for the Galactic disk 
(Galactocentric distance range of 4--11~kpc), namely 4.16. Based on the aforementioned ratios, the canonical C$^{18}$O and C$^{17}$O abundances were set to $4.1\times10^{-7}$ and $9.9\times10^{-8}$, respectively. With respect to the observed abundances, 
the CO depletion factors were derived to be $f_{\rm D}=27.3\pm1.8$ towards the 
core centre (C$^{18}$O data), and $f_{\rm D}=8.3\pm0.7$ in the envelope 
(C$^{17}$O data). The deuteration level and the CO depletion factors are given 
in the last two rows in Table~\ref{table:chemistry}. We note that the non-LTE 
analysis presented in Paper~III yielded a value of $f_{\rm D}=10.8\pm2.2$ towards
the core edge, i.e. a factor of $1.3\pm0.3$ times higher than the present value.

Assuming that the core mass we derived through SED fitting, 
$3.1\pm0.6$~M$_{\sun}$, is the mass within an effective radius, which corresponds to the size of the largest photometric 
aperture used, i.e. $R_{\rm eff}=19\farcs86$ or $\simeq0.04$~pc, the 
volume-averaged H$_2$ number density is estimated to be 
$\langle n({\rm H_2})\rangle=1.7\pm0.3\times10^5$~cm$^{-3}$ (see Eq.~(1) in 
Paper~III). Following the analysis presented in Miettinen (2012a, 
Sect.~5.5 therein), the CO depletion timescale at the aforementioned density (and adopting a $\delta_{\rm dg}$ ratio of 1/141) is estimated to be 
$\tau_{\rm dep}\sim3.4\pm0.6\times10^4$~yr. This can be 
interpreted as a lower limit to the age of SMM3. 

\begin{table}
\caption{Molecular column densities, fractional abundances with respect to 
H$_2$, and the degrees of deuteration and CO depletion.}
\small
\label{table:chemistry}
\begin{tabular}{c c c}
\hline\hline 
Species & $N$ [cm$^{-2}$] & $x$\\
\hline
NH$_3$ & $1.5\pm0.2\times10^{15}$ & $6.6\pm0.9\times10^{-8}$\\[1ex]
\textit{p}-H$_2$CO\tablenotemark{a} & $1.0\pm0.3\times10^{12}$ & $2.0\pm0.6\times10^{-11}$\\[1ex]
\textit{p}-H$_2$CO\tablenotemark{b} & $2.2\pm0.4\times10^{13}$ & $3.1\pm1.0\times10^{-9}$\\[1ex]
CH$_3$OH & $6.8\times10^{14}$\tablenotemark{c} & $9.4\pm2.5\times10^{-8}$\\[1ex]
C$^{18}$O & $7.1\pm0.8\times10^{14}$ & $1.5\pm0.1\times10^{-8}$\\[1ex]
SO & $8.1\pm1.2\times10^{12}$ & $1.6\pm0.2\times10^{-10}$\\[1ex]
C$^{17}$O & $3.2\pm0.4\times10^{14}$ & $1.2\pm0.1\times10^{-8}$\\[1ex]
N$_2$D$^+$ & $1.7\pm0.5\times10^{12}$ & $4.8\pm1.4\times10^{-11}$\\[1ex]
N$_2$H$^+$ & $1.2\pm0.4\times10^{13}$ & $2.9\pm0.9\times10^{-10}$\\[1ex]
DCO$^+$\tablenotemark{d} & $1.3\pm0.5\times10^{13}$ & $2.6\pm1.0\times10^{-10}$\\[1ex]
DCO$^+$\tablenotemark{e} & $6.2\pm2.9\times10^{11}$ & $2.0\pm0.9\times10^{-11}$\\[1ex]
\hline 
                         & Core centre & Envelope \\[1ex]  
[N$_2$D$^+$]/[N$_2$H$^+$] &   \ldots    & $0.14\pm0.06$ \\[1ex]
$f_{\rm D}({\rm CO})$ & $27.3\pm1.8$ & $8.3\pm0.7$ \\[1ex]
\hline 
\end{tabular} 
\tablenotetext{a}{The narrow-line component, which likely originates in \\ the quiescent 
envelope.}\tablenotetext{b}{The broad-line/warm component, which likely originates in \\ 
the outflow gas.}\tablenotetext{c}{The estimated error is unrealistically large  
(much larger \\ than the nominal value), and is therefore not reported.}\tablenotetext{d}{From the $J=3-2$ line observation towards the core centre.}\tablenotetext{e}{From the previous $J=4-3$ line observation towards the core \\ envelope.}   
\end{table}

\section{Discussion}

\subsection{Fragmentation and protostellar activity in SMM3}

Owing to the revised fundamental physical properties of SMM3, we are in a position to re-investigate its fragmentation characteristics. At a gas temperature of
$T_{\rm kin}=11.2\pm0.5$~K, the isothermal sound speed is 
$c_{\rm s}=197.5\pm4.4$~m~s$^{-1}$, where the mean molecular weight per free 
particle was set to $\mu_{\rm p}=2.37$. The aforementioned values can be used to calculate the thermal Jeans length 

\begin{equation}
\lambda_{\rm J}=\sqrt{\frac{\pi c_{\rm s}^2}{G \langle \rho \rangle}}\, ,
\end{equation}
where $G$ is the gravitational constant, the mean mass density is 
$\langle \rho \rangle=\mu_{\rm H_2}m_{\rm H}\langle n({\rm H_2})\rangle$, and 
$m_{\rm H}$ is the mass of the hydrogen atom. The resulting Jeans length, 
$\lambda_{\rm J}\simeq0.05$~pc, is a factor of 1.4 shorter than our previous 
estimate (0.07~pc; Paper~III), where the difference can be mainly attributed to the higher gas density derived here. We note that the uncertainty propagated from those of $T_{\rm kin}$ and $\langle n({\rm H_2})\rangle$ is only 1~mpc.

If we use the observed \textit{p}-NH$_3(1,\,1)$ linewidth as 
a measure of the non-thermal velocity 
dispersion, $\sigma_{\rm NT}$ ($=169.9\pm4.2$~m~s$^{-1}$), the effective sound 
speed becomes $c_{\rm eff}=(c_{\rm s}^2+\sigma_{\rm NT}^2)^{1/2}=260.5\pm1.6$~m~s$^{-1}$. The corresponding effective Jeans length is 
$\lambda_{\rm J}^{\rm eff}\simeq0.06$~pc. Although not much different from 
the purely thermal value, $\lambda_{\rm J}^{\rm eff}$ is in better agreement 
with the observed projected distances of SMM3b and 3c from the protostar 
position (0.07--0.10~pc). Hence, the parent core might have fragmented as a result of Jeans-type instabi\-lity with density perturbations in a self-gravitating fluid having both the thermal 
and non-thermal motions (we note that in Paper~III we suggested a pure thermal Jeans fragmentation scenario due to the aforementioned longer $\lambda_{\rm J}$ value). Because information in the core is transported at the sound speed (being it thermal or effective one), the fragmentation timescale is 
expected to be comparable to the crossing time, $\tau_{\rm cross}=R/c_{\rm eff}$, 
where $R=0.07-0.10$~pc. This is equal to 
$\tau_{\rm cross}\sim2.6-3.8\times10^5$~yr, which is up to an order of magnitude longer than the estimated nominal CO depletion timescale (Sect.~4.2.3).

The present SED analysis and the previous studies (see Sect.~1) suggest
that SMM3 is in the Class~0 phase of stellar evolution. Observational estimates 
of the Class~0 lifetime are about $\sim1\times10^5$~yr (\citealp{enoch2009}; 
\citealp{evans2009}; Maury et al. 2011). In agreement with observations, 
Offner \& Arce (2014) performed radiation-hydrodynamic simulations of 
protostellar evolution including outflows, and obtained Stage~0 lifetimes of 
$1.4-2.3\times10^5$~yr, where the Stage~0 represents a theoretical counterpart 
of the observational Class~0 classification. 
These observational and theoretical lifetime estimates are comparable to the 
fragmentation timescale of SMM3, which supports a scenario of the age of SMM3 being a few times $10^5$~yr. 

In the present paper, we have presented the first 
signatures of an outflow activity in SMM3. These are \textit{i)} the broad 
lines of \textit{p}-H$_2$CO and CH$_3$OH; \textit{ii)} the warm gas 
($64\pm15$ K) associated with the broad-line component; and \textit{iii)} the 
protrusion-like feature seen at 4.5~$\mu$m (Fig.~\ref{figure:images}, bottom 
right panel), which is likely related to the shock emission near the accreting 
protostar. Outflow activity reasserts the Class 0 evolutionary stage of 
SMM3 (e.g. \citealp{bontemps1996}). 

The 350~$\mu$m flux densities of the subcondensations SMM3b and 3c are 
$250\pm60$~mJy and $240\pm60$~mJy, respectively (Paper~III). Assuming that 
the dust temperature is that resulting from the SED of SMM 3 ($15.1\pm0.1$~K),
and adopting the same dust model as in Sect.~4.1, in which case the dust 
opacity per unit dust mass at 350~$\mu$m is 
$\kappa_{\rm 350\, \mu m}=7.84$~cm$^2$~g$^{-1}$, the condensation masses are only $\sim0.06\pm0.01$~M$_{\sun}$. If we instead use as $T_{\rm dust}$ the gas temperature derived from ammonia, the mass estimates will become about $0.16\pm0.05$~M$_{\sun}$, i.e. a factor of $2.7\pm0.9$ higher. As discussed in the case of the prestellar core Orion B9--SMM6 by Miettinen \& Offner (2013b), these types of very low-mass condensations are likely not able to collapse to form stars without any additional mass accretion. Instead, they could represent the precursors of 
substellar-mass objects or brown dwarfs (e.g. \citealp{lee2013}). Alternatively, 
if the condensations are gravitationally unbound structures, they could 
disperse away in the course of time, an issue that could be solved by 
high-resolution molecular line observations. Finally, 
mechanical feedback from the protostellar outflow could affect the future 
evolution of the condensations (cf.~the proto- and prestellar core 
system IRAS~05399-0121/SMM1 in Orion~B9; \citealp{miettinen2013a}).

\subsection{Chemical properties of SMM3}

\subsubsection{NH$_3$ and N$_2$H$^+$ abundances}

The fractional abundances of the N-bearing species NH$_3$ and N$_2$H$^+$ we 
derived are $6.6\pm0.9\times10^{-8}$ and $2.9\pm0.9\times10^{-10}$. 
The value of $x({\rm NH_3})$ in low-mass dense cores is typically found to be 
a few times $10^{-8}$ (e.g. \citealp{friesen2009}; \citealp{busquet2009}). 
Morgan et al. (2010) derived a mean $x({\rm NH_3})$ value of 
$2.6\times10^{-8}$ towards the protostars embedded in bright-rimmed clouds. 
Their sources might represent 
the sites of triggered star formation, and could therefore resemble the case 
of SMM3 -- a core that might have initially formed as a result of external 
feedback. More recently, Marka et al. (2012) found that the average NH$_3$ 
abundance in their sample of globules hosting Class 0 protostars is 
$3\times10^{-8}$ with respect to H$_2$.\footnote{The authors reported the 
abundances with respect to the total hydrogen column density, which is here assumed to be $N_{\rm H}=2N({\rm H_2})$.} Compared to the aforementioned reference studies, 
the ammonia abundance is SMM3 appears to be elevated by a factor of about two or more, although differences in the assumptions of dust properties should be borne in mind. The chemical modelling of the Class 0 sources performed by Marka et al. (2012), which included reactions taking place on dust grain surfaces, predicted that an NH$_3$ abundance exceeds $\sim10^{-8}$ after $10^5$~yr of evolution (see 
also \citealp{hilyblant2010} for a comparable result). This compares well with 
the fragmentation timescale in SMM3 estimated above. For their sample of 
low-mass protostellar cores, Caselli et al. (2002b) found a mean N$_2$H$^+$ 
abundance of $3\pm2\times10^{-10}$, which is very similar to the one we have 
derived for SMM3.

The [NH$_3$]/[N$_2$H$^+$] ratio in SMM3, derived from the corresponding column 
densities, is $125\pm45$. The abundance ratio between these two species is 
known to show different values in starless and star-forming objects. 
For example, Hotzel et al. (2004), who studied the dense cores 
B217 and L1262, both associated with Class~I protostars, found that 
the above ratio is $\sim140-190$ in the starless parts of the cores, but only 
about $\sim60-90$ towards the protostars. Our value, measured towards the 
outer edge of SMM3, lies in between these two ranges, and hence is consistent with the observed trend. A similar behaviour is seen in IRAS~20293+3952, a site of clustered star formation (\citealp{palau2007}), and 
clustered low-mass star-forming core Ophiuchus B (\citealp{friesen2010}). 
In contrast, for their sample of dense cores in Perseus, Johnstone et al. 
(2010) found that the \textit{p}-NH$_3$/N$_2$H$^+$ column density ratio is 
fairly similar in protostellar cores ($20\pm7$) and in prestellar cores 
($25\pm12$). Their ratios also appear to be lower than found in other sources 
(we note that the statistical equilibrium value of the NH$_3$ 
\textit{ortho}/\textit{para} ratio is unity; e.g. \citealp{umemoto1999}). 

The chemical reactions controlling the [NH$_3$]/[N$_2$H$^+$] ratio were 
summarised by Fontani et al. (2012; Appendix~A therein). In starless 
cores, the physical conditions are such that both the 
CO and N$_2$ molecules can be heavily depleted. If this is the 
case, N$_2$H$^+$ cannot be efficiently formed by the reaction between H$_3^+$ 
and N$_2$. On the other hand, this is counterbalanced by the fact that 
N$_2$H$^+$ cannot be destroyed by the gas-phase CO, although it would serve 
as a channel for the N$_2$ production (${\rm CO}+{\rm N_2H^+}\rightarrow {\rm HCO^+}+{\rm N_2}$). 
Instead, in a gas with strong CO depletion, N$_2$H$^+$ is destroyed by the 
dissociative electron recombination. The absence of N$_2$ also diminishes the 
production of N$^+$, the cations from which NH$_3$ 
is ultimately formed via the reaction NH$_4^++{\rm e}^-$. However, the other 
routes to N$^+$, namely ${\rm CN}+{\rm He}^+$ and ${\rm NH_2}+{\rm He}^+$, can 
still operate. We also note that H$_3^+$, which also cannot be destroyed by CO 
in the case of strong CO depletion, is a potential destruction agent of NH$_3$. 
However, the end product of the reaction ${\rm NH_3}+{\rm H_3^+}$ is 
NH$_4^+$, the precursor of NH$_3$. For these reasons, the NH$_3$ abundance 
can sustain at the level where the [NH$_3$]/[N$_2$H$^+$] ratio is higher in 
starless cores (strong depletion) than in the protostellar cores (weaker 
depletion). It should be noted that the study of the  
high-mass star-forming region AFGL 5142 by Busquet et al. (2011) showed that 
the [NH$_3$]/[N$_2$H$^+$] ratio behaves opposite to that in low-mass 
star-forming regions. The authors concluded that the higher ratio 
seen towards the hot core position is the result of a higher dust temperature, 
leading to the desorption of CO molecules from the grains mantles. As a result, the gas-phase CO can destroy the N$_2$H$^+$ molecules, which results in a higher 
[NH$_3$]/[N$_2$H$^+$] ratio. Because SMM3 shows evidence for quite a strong CO 
depletion of $f_{\rm D}=27.3\pm1.8$ towards the core centre, the chemical scheme described above is probably responsible for the 
much higher abundance of ammonia compared to N$_2$H$^+$.   

\subsubsection{Depletion and deuteration}

As mentioned above, the CO molecules in SMM3 appear to be quite heavily 
depleted towards the protostar position, while it becomes lower by a factor 
of $3.3\pm0.4$ towards the outer core edge. A caveat here is that the 
two depletion factors were derived from two different isotopologues, namely C$^{17}$O for the envelope zone, and C$^{18}$O towards the core centre. This brings into question the direct comparison of the two depletion factors. Indeed, although the critical densities of the detected CO 
isotopologue transitions are very similar, the C$^{18}$O linewidth is 
$1.4\pm0.3$ times greater than that of C$^{17}$O. Although this is not a significant discrepancy, the observed C$^{18}$O emission could originate in a
more turbulent parts of the core. 

For comparison, for their sample of 20 Class 0 protostellar cores, 
Emprechtinger et al. (2009) derived CO depletion factors of $0.3\pm0.09-4.4\pm1.0$. These are significantly lower than what we have derived for SMM3. 
The depletion factor in the outer edge of SMM3 we found is more remisnicent 
to those seen in low-mass starless cores (e.g. \citealp{bacmann2002}; 
\citealp{crapsi2005}), but the value towards the core's 24~$\mu$m peak position 
stands out as an exceptionally high. 

The deuterium fractionation of N$_2$H$^+$, or the N$_2$D$^+$/N$_2$H$^+$ column 
density ratio, is found to be $0.14\pm0.06$ towards the core edge. This lies 
midway between the values found by Roberts \& Millar (2007) for their sample 
of Class 0 protostars ($0.06\pm0.01-0.31\pm0.05$). Emprechtinger et al. (2009) 
found N$_2$D$^+$/N$_2$H$^+$ ratios in the range $<0.029-0.271\pm0.024$ with an 
average value of 0.097. Among their source sample, most objects had a 
deuteration level of $<0.1$, while 20\% of the sources showed values of $>0.15$.
With respect to these results, the deuterium fractionation in SMM3 appears to 
be at a rather typical level among Class 0 objects. For comparison, in 
low-mass starless cores the N$_2$D$^+$/N$_2$H$^+$ ratio can be several tens of 
percent (\citealp{crapsi2005}), while intermediate-mass Class 0-type protostars 
show values that are more than ten times lower than in SMM3 
(\citealp{alonso2010}). A visual inspection of Fig.~3 in Emprechtinger et al. 
(2009) suggests that for a N$_2$D$^+$/N$_2$H$^+$ ratio we have derived for 
SMM3, the dust temperature is expected to be $\lesssim25$~K. 
This is qualitatively consistent with a value of $15.1\pm0.1$~K 
we obtained from the MBB SED fit. On the other hand, the correlation in 
the middle panel of Fig.~4 in Emprechtinger et al. (2009; see also their 
Fig.~10) suggests that the CO depletion factor would be $\sim3$ at the 
deuteration level seen in SMM3, while our observed value in the envelope is 
$2.8\pm0.2$ times higher. The fact that CO molecules appear to be 
more heavily depleted towards the new line observation target position 
suggests that the degree of deuterium fractionation there is also higher. 
A possible manifestation of this is that the estimated DCO$^+$ abundance is 
higher by a factor of $13.0\pm7.7$ towards the core centre than towards the 
core edge, but this discrepancy could be partly caused by the different 
transitions used in the analysis ($J=3-2$ and $J=4-3$, respectively). 

Recently, Kang et al. (2015) derived a deuterium fractionation of formaldehyde 
in SMM3 (towards the core centre), and they found a HDCO/H$_2$CO ratio of $0.31\pm0.06$, which is the highest value among their sample of 15 Class~0 objects. This high deuteration level led the authors to conclude that SMM3 is in a very early stage of protostellar evolution.

\subsubsection{H$_2$CO, CH$_3$OH, and SO -- outflow chemistry in SMM3}

Besides the narrow ($\Delta v=0.42$~km~s$^{-1}$) component of the 
\textit{p}-H$_2$CO$(3_{0,\,3}-2_{0,\,2})$ line detected towards SMM3, this line 
also exhibits a much wider ($\Delta v=8.22$~km~s$^{-1}$) component with 
blue- and redshifted wing emission. The other two transitions of 
\textit{p}-H$_2$CO we detected, $(3_{2,\,1}-2_{2,\,0})$ and 
$(3_{2,\,2}-2_{2,\,1})$, are also broad, more than 10~km~s$^{-1}$ in FWHM, 
and exhibit wing emission. The methanol line we detected, 
with a FWHM of 10.98~km~s$^{-1}$, is also 
significantly broader than most of the lines we have detected. 
The similarity between the FWHMs of the methanol and formaldehyde lines suggests that they 
originate in a common gas component. The rotational temperature derived from 
the \textit{p}-H$_2$CO lines, $64\pm15$~K, is considerably higher than the 
dust temperature in the envelope and the gas temperature derived from ammonia. 
The large linewidths and the relatively warm gas temperature can be understood 
if a protostellar outflow has swept up and shock-heated the surrounding 
medium.

The H$_2$CO and CH$_3$OH molecules are organic species, and they can form 
on dust grain surfaces through a common CO hydrogenation reaction sequence 
(${\rm CO}\rightarrow {\rm HCO}\rightarrow {\rm H_2CO}\rightarrow {\rm CH_3O}$ 
or H$_3$CO or CH$_2$OH or H$_2$COH $\rightarrow {\rm CH_3OH}$; e.g. 
Watanabe \& Kouchi 2002; \citealp{hiraoka2002}; \citealp{fuchs2009}, and 
references therein). The intermediate compound, solid formaldehyde, and the 
end product, solid methanol, have both been detected in absorption towards 
low-mass young stellar objects (YSOs; Pontoppidan et al. 2003; \citealp{boogert2008}). A more recent study of solid-phase CH$_3$OH in low-mass YSOs by Bottinelli et al. (2010) suggests that much of the CH$_3$OH is in a CO-rich ice layer, 
which conforms to the aforementioned formation path. We note that H$_2$CO can also be formed in the gas phase (e.g. \citealp{kahane1984}; \citealp{federman1991}), and the narrow \textit{p}-H$_2$CO line we detected is likely tracing a 
quiescent gas not enriched by the chemical compounds formed on dust grains. The 
estimated \textit{p}-H$_2$CO abundance for this component is very low, only 
$2.0\pm0.6\times10^{-11}$. We note that the total H$_2$CO column density derived by Kang et al. (2015), $N({\rm H_2CO})=3.3\pm0.4\times10^{12}$~cm$^{-2}$ at $T_{\rm ex}=10$~K, is in good agreement with our \textit{p}-H$_2$CO column density if the \textit{ortho}/\textit{para} ratio is 3:1 as assumed by the authors. 

In contrast, the fractional \textit{p}-H$_2$CO abundance is found to be 
$155\pm70$ times higher for the broad component than for the narrow one. 
The origin of the H$_2$CO abundance 
enhancement in low-mass protostars can be understood in terms of the 
liberation of the ice mantles (\citealp{schoier2004}). We note that there are 
also gas-phase formation routes for CH$_3$OH, which start from the reaction 
between CH$_3^+$ and H$_2$O or between H$_3$CO$^+$ and H$_2$CO. The resulting 
protonated methanol, CH$_3$OH$_2^+$, can recombine with an electron and 
dissociate to produce CH$_3$OH (\citealp{garrod2006}; \citealp{geppert2006}). 
However, the gas-phase syntheses are not able to produce the high fractional 
abundances like observed here towards SMM3 ($9.4\pm2.5\times10^{-8}$). 
The thermal desorption of CH$_3$OH requires a dust temperature of 
at least $\sim80$~K (\citealp{brown2007}; \citealp{green2009}). Although highly 
uncertain, the H$_2$CO rotational temperature we derived does not suggest the 
dust temperature to be sufficiently high for CH$_3$OH molecules to 
sublimate. Hence, it seems possible that an outflow driven by SMM3 
has sputtered the icy grain mantles (in impacts with gas-phase H$_2$ and He) 
so that H$_2$CO and CH$_3$OH were released into the gas phase. On the other 
hand, the high CO depletion factors we derived suggest that the grain ices are 
rich in CO, and if CH$_3$OH molecules are embedded in CO-rich ice layers, their 
thermal evaporation temperature can be considerably lower ($\sim30$~K; see 
\citealp{maret2005}).

The \textit{p}-H$_2$CO/CH$_3$OH column density ratio for the broad line 
component is found to be $0.03\pm0.005$. This value represents a lower limit 
to the total H$_2$CO/CH$_3$OH ratio, which depends on the 
\textit{o}/\textit{p} ratio. Based on the observed abundances of both the \textit{ortho} and \textit{para} forms of H$_2$CO in low-mass dense cores, J{\o}rgensen et al. (2005) derived a \textit{o}/\textit{p} ratio of $1.6\pm0.3$. The authors interpreted this to be consistent with thermalisation at 10--15~K on 
dust grains. If we assume that the \textit{o}/\textit{p} ratio is $\simeq1.6$, 
we obtain a total H$_2$CO/CH$_3$OH column density ratio of $\simeq0.08\pm0.01$, while for a \textit{o}/\textit{p} ratio, which is equal to the 
relative statistical weights of 3:1, the total H$_2$CO/CH$_3$OH ratio becomes 
$0.13\pm0.02$. The H$_2$CO/CH$_3$OH ice abundance ratio 
in low-mass YSOs is found to be in the range $\sim0.2-6$ 
(\citealp{boogert2008}), which is higher than the gas-phase abundance ratio towards SMM3. Hence, it is possible that the ices are not completely sublimated into the gas phase. Interestingly, if the total H$_2$CO/CH$_3$OH ratio for SMM3 is 
$\sim0.1$, and $x({\rm CH_3OH})\sim10^{-7}$, then these properties would resemble those derived for the Galactic centre clouds 
where shocks (caused by expanding bubbles, cloud-cloud collisions, etc.) are 
believed to have ejected the species from the grain mantles 
(\citealp{requena2006}). In contrast, for the hot interiors of Class 0 sources, 
i.e. hot corinos, the H$_2$CO/CH$_3$OH ratio is found to be higher, 
in the range $>0.3-4.3$ (\citealp{maret2005}; their Table~3), 
which are comparable to the aforementioned ice abundance ratios. 
In hot corinos the dust temperature exceeds 100 K, and the evaporation of ice mantles is the result of 
radiative heating by the central protostar. Moreover, in the Horsehead photodissociation region (PDR) in Orion~B, the H$_2$CO/CH$_3$OH ratio is found to be $2.3\pm0.4$ (\citealp{guzman2013}). Guzm{\'a}n et al. (2013) concluded that in the 
UV-illuminated PDR both H$_2$CO and CH$_3$OH are released from the grain 
mantles through photodesorption.

The SO line we detected is narrow ($\Delta v=0.68$~km~s$^{-1}$), but 
low-intensity wing emission can be seen on both sides of it (see 
\citealp{codella2002} for similar spectra towards the CB34 globule, which harbours
a cluster of Class 0 objects). The derived fractional abundance of SO, 
$1.6\pm0.2\times10^{-10}$, is very low, as for example the average abundance 
derived by Buckle \& Fuller (2003) for their sample of Class~0 objects is 
$3.1\pm0.9\times10^{-9}$, and in the starless TMC-1 cloud the abundance is 
found to be $\sim10^{-8}$ (\citealp{lique2006}). 

While our narrow SO line is 
probably originating in the quiescent envelope, where SO is formed through the 
reactions ${\rm S}+{\rm OH}$ and ${\rm S}+{\rm O_2}$ (e.g. \citealp{turner1995}), 
the weak line wings provide a hint of an outflowing SO gas. The SO emission is 
indeed known to be a tracer of protostellar outflows 
(e.g. \citealp{chernin1994}; \citealp{lee2010}; \citealp{tafalla2013}). 
Outflow shocks can first release H$_2$S molecules from dust grains, 
and subsequent hydrogenation reactions produce HS 
molecules and S atoms (${\rm H_2S}+{\rm H}\rightarrow {\rm HS}+{\rm H_2}$; 
${\rm HS}+{\rm H}\rightarrow {\rm S}+{\rm H_2}$; \citealp{mitchell1984}; 
\citealp{charnley1997}). The oxidation reactions ${\rm HS}+{\rm O}$ and 
${\rm S}+{\rm O_2}$ can then lead to the formation of SO (see 
\citealp{bachiller2001}). For example, Lee et al. (2010) derived an SO abundance 
of $\sim2\times10^{-6}$ towards the HH211 jet driven by a Class 0 protostar, 
which shows that a significant abundance enhancement can take place in 
low-mass outflows. Some of the evolutionary models of the sulphur chemistry 
by Buckle \& Fuller (2003) suggest that, after $\sim10^5$ yr, 
the abundance of H$_2$S starts to drop, which leads to a rapid decrease in 
the SO abundance. This could explain the very weak SO wing emission seen 
towards SMM3, and agrees with the observational estimates of the 
Class 0 lifetime of about $\sim1\times10^5$ yr (e.g. \citealp{evans2009}; 
see also \citealp{offner2014} for a comparable result from simulations).
Interestingly, some of the Buckle \& Fuller (2003) models, for example the one 
with a gas temperature of 10~K, H$_2$ density of $10^5$~cm$^{-3}$, and a
cosmic-ray ionisation rate of $\zeta_{\rm H}=1.3\times10^{-16}$ s$^{-1}$, which is
ten times the standard $\zeta_{\rm H}$ (their Fig.~7, bottom left), predict SO 
abundances comparable to that observed in SMM3 (a few times $10^{-10}$) after 
$10^5$ yr, so perhaps the narrow-line component could also be (partly) tracing 
a gas component that was affected by outflows in the past

\section{Summary and conclusions}

We used the APEX telescope to carry out follow-up molecular line observations 
towards the protostellar core SMM3 , which is embedded in the filamentary 
Orion B9 star-forming region. The new data were used in conjunction with 
our earlier APEX data (including SABOCA and LABOCA continuum data), and NH$_3$ 
observations from the Effelsberg 100~m telescope. The main results are 
summarised as follows.

\begin{enumerate}
\item From the observed frequency range $\sim218.2-222.2$~GHz, the following 
chemical compounds were identified: $^{13}$CO, C$^{18}$O, SO, 
\textit{p}-H$_2$CO, and CH$_3$OH-E$_1$. The last two species play a key 
role in the synthesis of more complex organic molecules and prebiotic 
chemistry, which makes them particularly interesting compounds in the 
gas reservoir of a solar-type protostar like SMM3. 
Our new mapping observations of SMM3 were performed in the frequency range 
$\sim215.1-219.1$~GHz, from which DCO$^+(3-2)$ and 
\textit{p}-H$_2$CO$(3_{0,\,3}-2_{0,\,2})$ lines were identified.
\item Our revised SED analysis of SMM3 supports its Class 0 classification. 
The dust temperature, envelope mass, and luminosity 
were derived to be $15.1\pm0.1$~K, $3.1\pm0.6$~M$_{\sun}$, and $3.8\pm0.6$~L$_{\sun}$. 
The NH$_3$-based gas kinetic temperature was derived to be $T_{\rm kin}=11.2\pm0.5$~K. 
The revised analysis of the subfragments seen in our SABOCA 350~$\mu$m map suggests that SMM3 went 
through a Jeans-type fragmentation phase, where the initial density perturbations might have had contributions from both thermal and non-thermal motions.
\item The CO depletion factor derived from the new C$^{18}$O data towards the 
core centre is very high, $27.3\pm1.8$., while that
re-computed from our previous C$^{17}$O data towards the core edge is clearly 
lower, $8.3\pm0.7$. We also recalculated the degree of deuterium fractionation 
in the latter position, in terms of the N$_2$D$^+$/N$_2$H$^+$ ratio, and found a 
value of $0.14\pm0.06$. Even higher deuteration is to be expected towards the 
new line observation target position because of the stronger CO freeze 
out.
\item The new spectral-line mapping observations revealed that SMM3 is 
associated with extended DCO$^+$ and \textit{p}-H$_2$CO emission (as compared with the 350~$\mu$m-emitting region), and both the line emissions appear to be 
elongated in the east-west direction. Besides the systemic velocity of 
$\sim8.5$~km~s$^{-1}$, emission from \textit{p}-H$_2$CO$(3_{0,\,3}-2_{0,\,2})$ was also detected at a radial velocity of 1.5~km~s$^{-1}$, which 
concentrates to the east and northeast of SMM3, similarly to the spatial distributions of $^{13}$CO$(2-1)$ and C$^{18}$O$(2-1)$ seen earlier by Miettinen (2012b).
\item The single-pointing observations showed that the $3_{0,\,3}-2_{0,\,2}$ 
line of \textit{p}-H$_2$CO exhibits two components, a narrow one and a broad 
one. The other two \textit{p}-H$_2$CO lines we detected, $3_{2,\,1}-2_{2,\,0}$ 
and $3_{2,\,2}-2_{2,\,1}$, are also broad. Hence, a rotational diagram was 
constructed for the broad component of \textit{p}-H$_2$CO, which yielded a 
rotational temperature of $64\pm15$~K. The detected methanol line has a width 
comparable to those of the broad formaldehyde lines, and is hence likely  
tracing the same warm gas component.
\item We interpret the broad \textit{p}-H$_2$CO and CH$_3$OH lines, and the 
e\-levated gas temperature, to be the first clear evidence of shock processing 
and outflow activity in SMM3. The abundance of \textit{p}-H$_2$CO in the 
broad component is enhanced by two orders of magnitude with respect to the 
quiescent gas component. Additionally, the protrusion-like emission feature 
seen in the \textit{Spitzer} 4.5~$\mu$m image is likely related to shock emission.
\item The detected SO line shows a narrow component at the systemic velocity, 
and weak wings on both sides of it. The wing emission points towards a weak SO 
outflow, while the narrow component is probably tracing the quiescent envelope.
\item The estimated fragmentation timescale of SMM3, and the observed chemical 
characteristics all suggest that the age of SMM3 is a few times $10^5$~yr, 
in agreement with its inferred Class~0 evolutionary stage. A dedicated chemical modelling would be useful in setting tighter constraints on the source age.
\end{enumerate}

Putting the results from the previous studies and the present
one together, we are in a position to place SMM3 in the wider context of Class~0 objects. Stutz et al. 
(2013) classified SMM3 as a so-called PACS Bright Red source, or PBRs. This source population is composed of extreme, red Class~0 objects with presumably high-density envelopes and high mass infall rates, and the median values of their MBB-based dust temperature, envelope mass, luminosity, and $L_{\rm submm}/L$ ratio are 19.6~K, 0.6~M$_{\sun}$, 1.8~L$_{\sun}$, and 2.7\% (see Table~8 in S13).\footnote{We note that these median values were calculated by including the SMM3 values derived by S13, but if they are omitted, the median values are essentially the same. The median envelope mass reported here was scaled to the presently assumed dust-to-gas ratio.} Although the physical properties of SMM3 we have derived in the present work are more extreme than the typical PBRs' properties (it is colder, more massive, and more luminous), it can still be classified as a PBRs in agreement with S13 because this population was also found to contain sources with properties comparable to those we have derived. We note that the Orion~B cloud appears to contain a relatively high fraction of PBRs-type objects (17\% of the known protostars in Orion~B) compared to that in Orion~A (1\%; S13).

We can also draw a conclusion that SMM3 exhibits a rich chemistry. 
It is possible that this Class~0 protostellar core hosts a so-called hot corino where the gas-phase chemistry can be as rich as in the hot molecular cores associated with high-mass star formation. This can be tested through high resolution 
interferometric multi-line observations. Such observations would also be useful 
to examine whether SMM3 drives a chemically rich/active molecular outflow, as our detection of the broad formaldehyde and methanol lines already suggest. In a more general context of low-mass star formation, SMM3 has the potential to become a useful target source of chemical evolution in a triggered star-forming region (feedback from NGC~2024, which could be ultimately linked to the nearby Ori OB1 association within the Ori-Eri superbubble). 
By comparing its properties with those of Class~0 objects in more isolated, quiescent regions, 
it could be possible to investigate whether its (chemical) evolution could have been accelerated 
as a result of more dynamic environment. The observed fragmentation of the SMM3 core indeed suggests 
that it has had a dynamical history, and is a fairly atypical object compared to the general Class~0 population in the Galaxy.

%
%

%

%
%

%

%
\acknowledgments

I would like to thank the referee for providing helpful, constructive 
comments and suggestions that improved the content of this paper. 
This publication is based on data acquired with the Atacama Pathfinder 
EXperiment (APEX) under programmes {\tt 079.F-9313(A)}, {\tt 084.F-9304(A)}, 
{\tt 084.F-9312(A)}, {\tt 092.F-9313(A)}, and {\tt 092.F-9314(A)}. 
APEX is a collaboration between the Max-Planck-Institut f\"{u}r 
Radioastronomie, the European Southern Observatory, and the Onsala Space 
Observatory. I would like to thank the staff at the APEX telescope for 
performing the service-mode heterodyne and bolometer observations presented 
in this paper. The research for this paper was financially supported by 
the Academy of Finland, grant no.~132291. This research has made use of 
NASA's Astrophysics Data System, and the NASA/IPAC Infrared Science Archive, 
which is operated by the JPL, California Institute of Technology, 
under contract with the NASA. This study also made use of APLpy, 
an open-source plotting package for Python hosted at 
{\tt http://aplpy.github.com}.


%
\bibliographystyle{plainnat}


%

\end{document}